\documentclass[11pt,a4paper]{JHEP3}
\def\dirac#1{#1\llap{/}}
\newcommand{\dint}[2]{\int \frac{d^#1#2}{(2\pi)^#1}}
\usepackage{amsmath}

\usepackage{geometry}
\usepackage{cite}

\usepackage{graphicx,axodraw}

\title{Three-particle contributions to the
renormalisation\\[0.5em]
of $B$-meson light-cone distribution amplitudes}

\author{S.~Descotes-Genon\\Laboratoire de Physique Th\'eorique
CNRS/Univ. Paris-Sud 11, F-91405 Orsay, France\\
         E-mail:\email{sebastien.descotes-genon@th.u-psud.fr}}
\author{N.~Offen\\
Laboratoire de Physique Th\'eorique
CNRS/Univ. Paris-Sud 11, F-91405 Orsay, France\\
        E-mail: \email{nils.offen@th.u-psud.fr}}

\abstract{We study light-cone distribution amplitudes of heavy-light systems, such as a
$B$-meson. By an explicit computation, we determine how two-parton
distribution amplitudes mix with three-parton ones at one loop: $\phi_+$ is
shown to mix only into itself, whereas $\phi_-$ mixes with the difference of
three-parton distribution amplitudes $\Psi_A-\Psi_V$. We determine the corresponding
anomalous dimension and we check the gauge independence of
our result by considering a general covariant gauge. Finally, we comment on some
implications of our result for phenomenological models of these distribution amplitudes.
} 
\keywords{B-physics, QCD, Renormalization Group}
\begin{document}







%
%
%




\section{Introduction}

$B$ physics provides outstanding opportunities to test the Standard Model. At the end of the
era of $B$-factories, a remarkable sample of processes has been measured accurately. On the theoretical side, the dynamics of heavy-light systems is naturally dominated by two scales : the heavy-quark mass $m_b$ corresponding to hard contributions and a hadronic scale $\Lambda$ related to soft, hadronic, contributions. It is therefore natural to deal with $B$-mesons in the framework of the heavy-quark expansion, expanding observables in powers of $1/m_b$, and to attempt to factorise hard and soft dynamics.

This has been progressively achieved  within the frameworks of QCD
factorisation~\cite{Beneke:2000ry,Beneke:2001ev,Beneke:2003zv} and Soft-Collinear Effective Theory~\cite{Bauer:2000yr,Bauer:2001yt,Bauer:2005kd,Beneke:2002ni,Beneke:2002ph}. The
hadronic  input collecting soft physics are not only the form factors, but also the 
light-cone distribution amplitudes, defined generally as the matrix element of a non-local operator along a light-like direction.
This quantity provides the amplitude of probability of finding partons inside
a hadron with given fractions of the hadron momentum. For light hadrons, one
can define a twist, corresponding to the power of the hard scale to which the
distribution amplitude will contribute to a given process, which is related to
the number of partons and the Lorentz structure of interest~\cite{Braun:2003rp}.
 For
instance, the leading-twist distribution amplitude of the pion can be defined as:
\begin{equation}
\langle \pi(p)|\bar{q}(z_2)[z_2,z_1]\gamma_\mu\gamma_5 q(z_1)|0\rangle =
  - i f_\pi p_\mu \int_0^1 dx\, e^{i(xp\cdot z_2+\bar{x}p\cdot z_1)} \phi_\pi(x)
\end{equation}
where $(z_2-z_1)^2=0$, $\bar{x}=1-x$, and the path-ordered exponential,
ensuring the gauge invariance of the expression, reads:
\begin{equation}
[z_2,z_1]=P\exp\left[ig_s\int_{z_1}^{z_2} dy_\mu A^\mu(y)\right]
\end{equation}
This distribution amplitude corresponds to picking up a pair of valence quark
and anti-quark in the pion, carrying a fraction $x$ and $\bar{x}$ of the pion
momentum respectively.

In the early days of QCD and the parton model, distribution amplitudes of
light mesons were identified as key tools to analyse exclusive processes at
high energy through factorisation, and they have been analysed extensively~\cite{Lepage:1980fj,Efremov:1979qk,Chernyak:1983ej}. Phenomenologically, sum
rules and lattice have been used to determine the values of their lowest
moments~\cite{Braun:1997kw,Baron:2007ti,Braun:2007zr,Boyle:2008nj}.  Mathematically, the structure of the distribution amplitudes was described by exploiting properties of the
conformal group, providing a geometrical definition of twist compatible with
the phenomenological one in the case of light mesons~\cite{Braun:2003rp}.

The case of heavy-light mesons has been discussed only recently~\cite{Grozin:2005iz,Grozin:1996pq}, mainly due to
their importance in relation with $B$-physics, as shown in the case of non-leptonic decays~\cite{Beneke:2000ry,Beneke:2001ev,Beneke:2003zv},
semileptonic decays~\cite{Beneke:2003pa,Beneke:2003xr}, radiative decays~\cite{DescotesGenon:2002mw,DescotesGenon:2002ja,DescotesGenon:2004hd,Beneke:2001at,Beneke:2000wa}. It is possible to define
two two-parton distribution amplitudes  $\phi_+$ and $\phi_-$ from the
most simple non-local matrix element:
\begin{equation}
\langle 0|\bar{q}_\beta(z)[z,0] (h_v)_\alpha(0)|B(p)\rangle =
-i\frac{\hat{f}_B(\mu)}{4} 
  \left[(1+\dirac{v})\left(\tilde{\phi}_+(t)+\frac{\dirac{z}}{2t}[\tilde\phi_-(t)-\tilde\phi_+(t)] \right)\gamma_5\right]_{\alpha\beta}
\end{equation}
with the usual definitions of Heavy Quark Effective Theory (HQET) for the velocity $v=p/M_B$, the heavy-quark
projection $h_v$ and the decay constant $\hat{f}_B$. The Fourier transforms of the distribution amplitudes,
depending on $t=v\cdot z$ read:
\begin{equation}
\tilde\phi_\pm(t)=\int_0^\infty d\omega\ e^{-i\omega t}\phi_\pm(\omega).
\end{equation}
It turns out that only one distribution amplitude, $\phi_+$, enters most of the computations considered in the
framework of factorisation (non-leptonic decays, $B\to V\gamma$, $B\to
\gamma\ell\nu$) at leading order in $1/m_B$. In cases where at least one of the outgoing particles has
not a light-like momentum, factorisation may still
hold, but the formula involves the two distribution amplitudes of the $B$
meson. This is in particular the case for  $B\to V\gamma^*$, which has an
important potential to test the Standard Model~\cite{Burdman:1998mk,Ali:1999mm}.

Another interesting place where the distribution amplitudes naturally occur is
light-cone sum rules. These sum rules allow in particular for a determination
of form factors of phenomenological relevance such as $B\to \pi$, $B\to
\rho$. The purpose of sum rules is to reexpress those form factors in terms of
an integral of the distribution amplitude of one of the external mesons with a
kernel resulting from the expansion of a properly chosen correlator along the
light-cone. Recently, sum rules with interesting properties have been proposed to relate these form
factors with $B$ meson distribution amplitudes. In this context, $\phi_-$
plays a dominant role and its modeling can improve the determination of the
form factors~\cite{Khodjamirian:2005ea,DeFazio:2007hw,DeFazio:2005dx,Khodjamirian:2006st}.

One can also define distribution amplitudes beyond the two-parton
level. Interestingly, one can exploit quark equations of motion in order to
relate two-parton and three-parton distribution amplitudes, which are defined
in the case of heavy-light mesons through:
\begin{eqnarray}
&&\langle 0|\bar{q}_\beta(z)[z,uz] gG_{\mu\nu}(uz)z^\nu [uz,0]
(h_v)_\alpha(0)|B(p)\rangle\\ \nonumber
&&\qquad\qquad =\frac{\hat{f}_B(\mu) M}{4}
  \left[(1+\dirac{v})
  \left[(v_\mu \dirac{z} -
    t\gamma_\mu)\left(\tilde\Psi_A(t,u)-\tilde\Psi_V(t,u)\right)
    -i\sigma_{\mu\nu}z^\nu \tilde\Psi_V(t,u)\right.\right.\\ \nonumber
&&\left.\left.\qquad\qquad\qquad\qquad\qquad
 - z_\mu \tilde{X}_A(t,u)
    +\frac{z_\mu \dirac{z}}{t} \tilde{Y}_A(t,u)
   \right]\gamma_5\right]_{\alpha\beta}
\end{eqnarray}

A good model for the distribution amplitudes of $\phi_+$ and $\phi_-$ must
embed as many theoretical constraints as possible. Not much can be said for
sure, apart from sum rule estimates~\cite{Ball:2003fq,Braun:2003wx}, the expected behaviour at the origin 
($\phi_+(\omega)\sim\omega$ and $\phi_-(\omega)\sim 1$ for $\omega\to 0$),
and the properties of these quantities under renormalisation, which can be
derived from perturbation theory.
One must notice that differences between light-meson and heavy-light
meson in terms of the renormalisation properties~\cite{Braun:2003wx} exist: the limit in one case is
the chiral limit, which modifies only long-distance properties, whereas the
heavy-quark limit affects short-distance features of the theory (the UV
structure of HQET is qualitatively different from QCD).
This explains the non-commutation of the heavy-quark limit with the light-cone
limit (contrary to the chiral limit). In the case of heavy-quark distribution amplitude, there is no equivalent of
Gegenbauer moments of light-meson distribution amplitudes, which mix only into themselves
under renormalisation.

The RGE behaviour of $\phi_+$ and $\phi_-$ has been investigated in
refs.~\cite{Lange:2003ff} and~\cite{Bell:2008er}. 
It was shown in particular that RGE generates a radiative tail, leading to a
divergence of positive moments of these quantities. In particular, there is no
absolute normalisation of $\phi_\pm$ to 1 from its zeroth moment, so that a naive partonic interpretation like in the pion case is not possible after renormalisation.
To obtain the first inverse moments of
$\phi_+$ and $\phi_-$, which are relevant phenomenologically, one needs the knowledge of the whole distribution
amplitude. But the previous behaviours and models were derived in the
two-parton approximation, even though the equation of motions indicate the
potential mixing with three-parton distribution amplitudes.

The goal of this paper is to understand the renormalisation of these objects at the first nontrivial order of
the strong coupling constant, including the contribution from three-parton distribution amplitudes. This task requires us to consider a three-parton external state (a quark, an antiquark and a gluon).
In Sec.~2, we recall the behaviour of $\phi_+$ and $\phi_-$ under
renormalisation as determined from a two-parton external state. In Sec.~3, we
give the one-loop diagrams and their ultraviolet divergences 
in the case of a three-parton external state. In Sec.~4, we determine the
mixing of $\phi_+$ and $\phi_-$ with three-parton distribution amplitudes at
one loop, and we show that there is no mixing in the case of $\phi_+$, whereas $\phi_-$ mixes
with the difference $\Psi_A-\Psi_V$. In Sec.~4, we extend our calculation,
performed in the Feynman gauge, to a general covariant gauge, which provides a
check of the gauge independence of our results (more detailed results are
given in App.~A). In Sec.~5, we make a few comments before concluding.

\section{Two-parton $B$-meson distribution amplitudes}\label{sec:twopart}

We set up our framework by introducing notation and definitions, and we recall
the results obtained on the renormalisation of the $B$-meson distribution amplitudes.
We define light-cone directions by two vectors
\begin{equation}
n_+^2=n_-^2=0 \qquad n_+\cdot n_-=2 \qquad v=(n_++n_-)/2
\end{equation}
so that an arbitrary vector can be projected as
\begin{equation}
q_\mu=(n_+\cdot q) \frac{n_{-,\mu}}{2} + (n_-\cdot q) \frac{n_{+,\mu}}{2} + q_{\perp\mu}
     =q_+ \frac{n_{-,\mu}}{2} + q_- \frac{n_{+,\mu}}{2} + q_{\perp\mu}
\end{equation}

The computation of the renormalisation properties of the distribution
amplitudes requires us to consider matrix elements of the relevant operators
\begin{eqnarray}
O_+^H(\omega)&=&
  \frac{1}{2\pi}\int dt  e^{i\omega t }
    \langle 0 | \bar{q}(z) [z,0] \dirac{n}_+ \Gamma h_v(0) |H \rangle \\
O_-^H(\omega)&=&
  \frac{1}{2\pi}\int dt  e^{i\omega t }
    \langle 0 | \bar{q}(z) [z,0] \dirac{n}_- \Gamma h_v(0) |H \rangle \\
O_3^H(\omega,\xi)&=&
  \frac{1}{(2\pi)^2}\int dt  e^{i\omega t }\int du e^{i\xi ut } 
    \langle 0 | \bar{q}(z) [z,uz] g_s G_{\mu\nu}(uz) z^\nu [uz,0] \Gamma 
         h_v(0) |H \rangle 
\end{eqnarray}
with $z$ parallel to $n_+$, i.e. $z_\mu=t  n_{+,\mu}$, 
$t =v\cdot z=z_-/2$ and the path-ordered exponential in 
the $n_+$ direction:
\begin{eqnarray} 
[z,0]&=&P\exp\left[i g_s \int_0^z dy_\mu A^\mu(y)\right]\\
 &=&1+ i g_s \int_0^1 d\alpha\ z_\mu A^\mu(\alpha z)
   -g_s^2 \int_0^1 d\alpha \int_0^\alpha d\beta z_\mu\ z_\nu\ 
           A^\mu(\alpha z)\ A^\nu(\beta z) + \ldots
\end{eqnarray}
We define the different distribution amplitudes in momentum space through
their Fourier transforms:
\begin{equation}
\phi_\pm(\omega)= \frac{1}{2\pi}\int dt e^{i\omega t} \tilde\phi_\pm(t)\qquad
F(\omega,\xi)= \frac{1}{(2\pi)^2}\int dt\int du t e^{i(\omega+u\xi)t} \tilde{F}(t,u)
\end{equation}
where $F=\Psi_V,\Psi_A, X_A, Y_A$.

The renormalisation group equation of $\phi_+$ and $\phi_-$ can be determined
by computing the mixing terms $Z$ defined in the following expression (since we work in the chiral limit, we omit mixing terms between $\phi_+$ and $\phi_-$ which were shown in ref.~\cite{Bell:2008er} to be proportional to the mass of the light-quark.)
\begin{equation}\label{eq:Zmix}
O_\pm^{H,ren}(\omega,\mu)=\int d\omega' Z^{-1}_\pm(\omega,\omega';\mu)
  O_\pm^{H,bare}(\omega') + \int d\omega'd\xi' Z_{\pm,3}^{-1}(\omega,\omega',\xi';\mu)
  O_3^{H,bare}(\omega',\xi')+\ldots
\end{equation}
where the ellipsis involves matrix elements of operators corresponding to a
higher number of partons (a similar equation for operators related to a higher number of partons instead of $O_\pm$ could be written, with operators corresponding to an arbitrary number of partons on the right-hand side). The behaviour under renormalisation being
a short-distance property, any choice of $H$ is allowed in principle.
For instance, $Z_\pm$ were computed in 
refs.~\cite{Lange:2003ff,Bell:2008er} using a two-parton external state
with on-shell quarks $H=h(p)\bar{q}(k)$,
for which one obtains at leading order:
\begin{equation}
O_\pm^H(\omega)=\delta(\omega-k_+) 
  \ \bar{v}\dirac{n}_\pm\Gamma u \qquad O_3^H(\omega,\omega')=0
\end{equation}
Computing NLO terms provides $Z_\pm$ in eq.~(\ref{eq:Zmix}).
However convenient, this choice of external state prevents us from determining
$Z_{\pm,3}$ describing the mixing between $O_\pm$ and
$O_3$, since $O_3^H$ vanishes then.

One can still write down the RGE restricted to $\phi_\pm$ as
\begin{eqnarray}
\frac{d}{d\log \mu}\phi_\pm(\omega,\mu)
  &=&-\int_0^\infty d\omega' \gamma_\pm(\omega,\omega',\mu) \phi_\pm(\omega',\mu)
  + \ldots\\
\gamma_\pm(\omega,\omega',\mu)&=&-\int d\tilde\omega 
   \frac{dZ_\pm^{-1} (\omega,\tilde\omega,\mu)}{d\log\mu}
   Z_\pm(\tilde\omega,\omega',\mu)
   -\gamma_F(\alpha_s) \delta(\omega-\omega')
\end{eqnarray}
where $\gamma_F$ is due to the normalisation of the distribution amplitudes
which involves the (renormalisation-scale dependent) HQET decay constant.

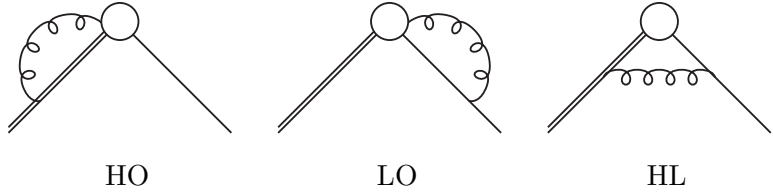
\begin{figure}
\begin{center}
\begin{tabular}{ccc}
\begin{picture}(90,50)(0,-5)
\SetScale{0.7}
\SetWidth{1}
\Line(0,3)(60,63)
\Line(0,0)(60,60)
\Line(60,60)(120,0)
\GlueArc(35,35)(25,45,225){4}{4}
\CCirc(60,60){10}{Black}{White}
\end{picture} &
\begin{picture}(90,50)(0,-5)
\SetScale{0.7}
\SetWidth{1}
\Line(0,3)(60,63)
\Line(0,0)(60,60)
\Line(60,60)(120,0)
\GlueArc(85,35)(25,-45,135){4}{4}
\CCirc(60,60){10}{Black}{White}
 
\end{picture}&
\begin{picture}(90,50)(0,-5)
\SetScale{0.7}
\SetWidth{1}
\Line(0,3)(60,63)
\Line(0,0)(60,60)
\Line(60,60)(120,0)

\CCirc(60,60){10}{Black}{White}
\Gluon(30,30)(90,30){4}{4}
\end{picture} \\
HO & LO & HL
\end{tabular}
\end{center}
\caption{Diagrams evaluated to determine the renormalisation constant $Z_\pm$ using a two-parton external state (the two letters describe the ends of the gluon line: Heavy quark, Light
quark or Operator, i.e. the path-ordered exponential).}
\label{fig:twopart}
\end{figure}

Working in dimensional regularisation with $d=4-2\varepsilon$ dimensions, in 
the Feynman gauge and in the $\bar{\rm MS}$ scheme, 
one obtains from the three diagrams shown in fig.~\ref{fig:twopart}~\cite{Lange:2003ff,Bell:2008er}:
\begin{eqnarray}
\left.Z_\pm\right|_{HO} &=& \frac{\alpha_s C_F}{4\pi}\times 2 
   \times \frac{1}{\varepsilon}\int_0^\infty \frac{dl_+}{l_+}
   \left(\frac{l_+^2}{\mu^2}\right)^{-\varepsilon}
     [\delta(\omega-\omega'-l_+)-\delta(\omega-\omega')] \label{eq:zpmho}\\
\left.Z_\pm\right|_{LO} &=& \frac{\alpha_s C_F}{4\pi} 
   \times (-2) \times \frac{1}{\varepsilon}\int_0^{\omega'} \frac{dl_+}{l_+}
   \frac{l_+-\omega'}{\omega'}
     [\delta(\omega'-\omega-l_+)-\delta(\omega'-\omega)]\\
\left.Z_-\right|_{HL} &=& \frac{\alpha_s C_F}{4\pi}\times 2 
   \times \frac{1}{\varepsilon}\int_0^{\omega'} \frac{dl_+}{\omega'}
   \delta(\omega'-\omega-l_+) \qquad\qquad
\left.Z_+\right|_{HL} =0
\end{eqnarray}
Only the diagram where the gluon line connects the two
external legs differs for $\phi_+$ and $\phi_-$.
In addition, one has contributions from the wave-function
renormalisation of the external legs:
\begin{eqnarray}\label{eq:zpmwfrh}
\left.Z_\pm\right|_{wfr-H} &=&  Z_h^{1/2} = \frac{\alpha_s C_F}{4\pi} 
   \times \frac{1}{\varepsilon}\delta(\omega-\omega')\\
\left.Z_\pm\right|_{wfr-L} &=&  Z_q^{1/2} = \frac{\alpha_s C_F}{4\pi}\times
\left(-\frac{1}{2}\right)
   \times \frac{1}{\varepsilon}\delta(\omega-\omega')\label{eq:zpmwfrl}
\end{eqnarray}
so that $Z_\pm$ is 
$\delta(\omega-\omega')$ corrected at one loop by the sum of eqs.~(\ref{eq:zpmho})-(\ref{eq:zpmwfrl}).

This leads to the following anomalous dimensions:
\begin{eqnarray}
\gamma_+^{(1)}&=&
  \left(\Gamma^{(1)}_{\rm cusp}\log\frac\mu\omega+\gamma^{(1)} \right)\delta(\omega-\omega')
 -\Gamma^{(1)}_{\rm cusp}\omega
    \left(\frac{\theta(\omega'-\omega)}{\omega'(\omega'-\omega)}
            +\frac{\theta(\omega-\omega')}{\omega(\omega-\omega')}\right)_+
        \\
\gamma_-^{(1)}&=&\gamma_+^{(1)}-\Gamma^{(1)}_{\rm cusp}\frac{\theta(\omega'-\omega)}{\omega'}
\end{eqnarray}
with
\begin{equation}
\Gamma^{(1)}_{\rm cusp}=4C_F \qquad \gamma^{(1)}=-2C_F \qquad \gamma_F^{(1)}=-3C_F
\end{equation}
Quantities with a superscript $(1)$ must be multiplied by $\frac{\alpha_s}{4\pi}$. The anomalous dimension is of the Sudakov type, which is
related to the fact that the operators of interest can be seen as containing
two Wilson lines, one from the heavy quark along the $v$ direction
(representing the interaction of soft gluons with $h_v$) from $-\infty$ to 0,
linked with another one along the $n_+$ direction from $0$ to $z$. The presence
of a cusp between the two Wilson lines is responsible for the appearance of a
Sudakov-like behaviour of the anomalous dimension~\cite{Korchemsky:1987wg,Korchemskaya:1992je,Lange:2003ff}.

\section{One-loop computation for three-particle external state}

We perform the same computation, taking as an external state
$H=h(p)g(\epsilon,q)\bar{q}(k)$ containing three partons. We compute the
diagrams at one loop, using dimensional regularisation, and we pay a special attention to 
separating $\varepsilon$-poles related to UV divergences and IR divergences carefully. 
We then identify the $\epsilon$-poles corresponding to UV divergences as part of the
renormalisation function in:
\begin{equation}
O_\pm^{H,bare}(\omega)=\int d\omega' Z_\pm(\omega,\omega';\mu)
  O_\pm^{H,ren}(\omega',\mu)+\int d\omega'd\xi' 
  Z_{\pm,3}(\omega,\omega',\xi';\mu) O_3^{H,ren}(\omega',\xi',\mu)
\end{equation}
Subtracting the contribution from  
  $Z_\pm \otimes O_\pm^H$ will yield the mixing term $Z_{\pm,3}$ between
  two and three-particle distribution amplitudes. 

\begin{figure}
\begin{center}
\begin{tabular}{c}
\begin{picture}(120,75)(0,-10)
\SetWidth{1}
\Line(0,3)(60,58)
\Line(0,0)(60,55)
\ArrowLine(30,26)(35,31)
\ArrowLine(60,55)(120,0)
\Text(20,28)[]{$p$}
\LongArrow(110,15)(90,35)
\Text(110,28)[]{$k$}
\Gluon(60,0)(60,45){4}{4}
\LongArrow(70,0)(70,15)
\Text(90,5)[]{$\mu,\epsilon,a$}
\CCirc(60,55){10}{Black}{White}
\Text(60,55)[]{$3$}
\end{picture}\\
$\displaystyle
A_{3\mu} : ig_s(q_+ \epsilon_\mu - q_\mu \epsilon_+)\ \bar{v}\Gamma T^a u\delta(\omega-k_+)\delta(\xi-q_+)
$
\end{tabular}
\end{center}
\caption{Leading-order contribution to the matrix element of $O_{3\mu}$ with a three-parton external state.}\label{fig:LOO3}
\end{figure}
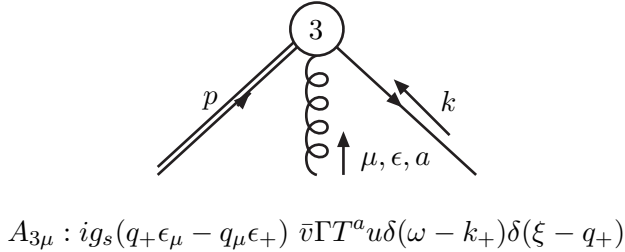

At leading order, we obtain three different contributions, denoted $A,B,C$, for the matrix
element of $O_\pm$, shown in fig.~\ref{fig:LO3part}.
The expression for $C$ can be simplified using the following relations
\begin{equation}
\epsilon\cdot q=0 \qquad \dirac{v} u=u \qquad \bar{v}\dirac{k}=0
\end{equation}
For $O_3$, we have a leading-order expression indicated in fig.~\ref{fig:LOO3}.

\begin{figure}
\begin{center}
\begin{picture}(120,60)(0,-10)
\SetWidth{1}
\Line(0,3)(60,58)
\Line(0,0)(60,55)
\ArrowLine(30,26)(35,31)
\ArrowLine(60,55)(120,0)
\Text(20,28)[]{$p$}
\LongArrow(110,15)(90,35)
\Text(110,28)[]{$k$}
\Gluon(60,0)(60,45){4}{4}
\LongArrow(70,0)(70,15)
\Text(90,5)[]{$\mu,\epsilon,a$}
\CCirc(60,55){10}{Black}{White}
\end{picture}

$\displaystyle
A : -g_s\frac{\epsilon_+}{q_+}
\left[\delta\left(\omega-k_+-q_+\right)-\delta\left(\omega-k_+\right)\right]
\bar{v}\dirac{n}_-\Gamma T^a u
$
\end{center}

\begin{center}
\begin{tabular}{ccc}
\begin{picture}(120,60)(0,-10)
\SetWidth{1}
\Line(0,3)(60,58)
\Line(0,0)(60,55)
\Line(60,55)(120,0)
\Gluon(60,0)(30,25){4}{4}
\CCirc(60,55){10}{Black}{White}
\end{picture}
&&
\begin{picture}(120,60)(0,-10)
\SetWidth{1}
\Line(0,3)(60,58)
\Line(0,0)(60,55)
\Line(60,55)(120,0)
\Gluon(60,0)(90,25){4}{4}
\CCirc(60,55){10}{Black}{White}
\end{picture}\\
$\displaystyle
B :- g_s\frac{v\cdot \epsilon}{v\cdot q}
\delta\left(\omega-k_+\right)
\ \bar{v}\dirac{n}_-\Gamma T^a u
$
&&
$\displaystyle
C : g_s\frac{1}{(k+q)^2}
\delta\left(\omega-k_+-q_+\right)
\ \bar{v}\dirac{\epsilon}(\dirac{k}+\dirac{q})\dirac{n}_-\Gamma T^a u
$
\end{tabular}
\end{center}
\caption{The three leading-order contributions to the matrix element of $O_\pm$ with a three-parton external state.}\label{fig:LO3part}
\end{figure}
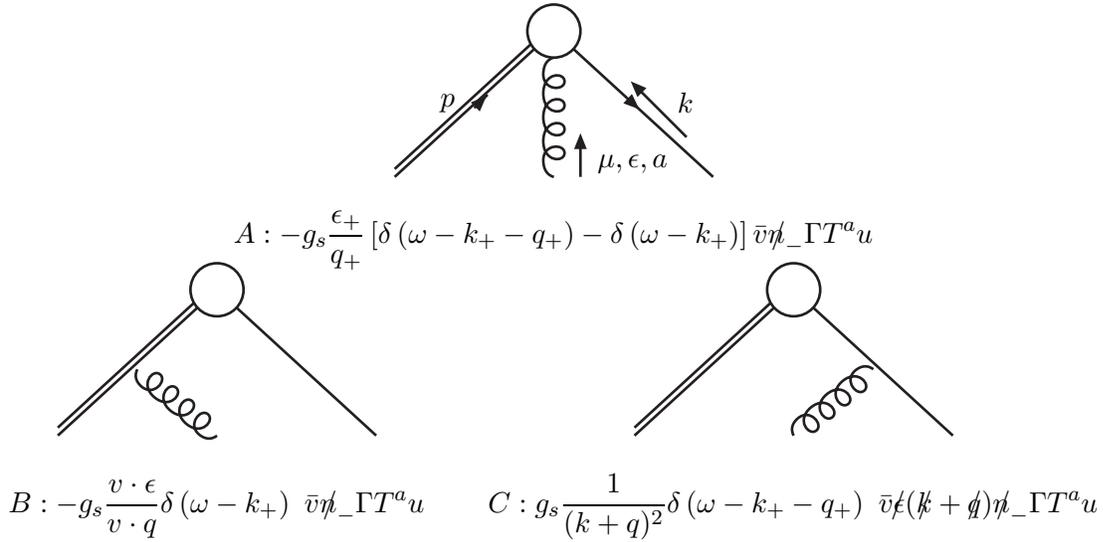

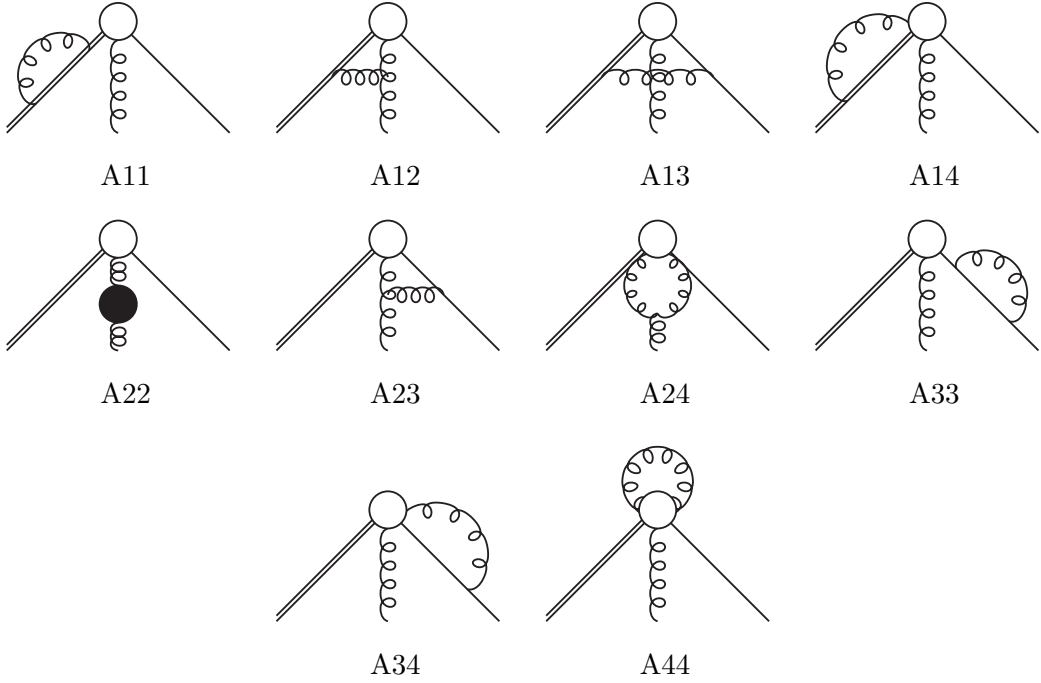
\begin{figure}
\begin{center}
\begin{tabular}{cccc}
\begin{picture}(90,50)(0,-5)
\SetScale{0.7}
\SetWidth{1}
\Line(0,3)(60,63)
\Line(0,0)(60,60)
\Line(60,60)(120,0)
\Gluon(60,0)(60,50){4}{4}
\CCirc(60,60){10}{Black}{White}
 
\GlueArc(30,30)(20,45,225){4}{4}
\end{picture} &
\begin{picture}(90,50)(0,-5)
\SetScale{0.7}
\SetWidth{1}
\Line(0,3)(60,63)
\Line(0,0)(60,60)
\Line(60,60)(120,0)
\Gluon(60,0)(60,50){4}{4}
\CCirc(60,60){10}{Black}{White}
 
\Gluon(30,30)(60,30){4}{3}
\end{picture} &
\begin{picture}(90,50)(0,-5)
\SetScale{0.7}
\SetWidth{1}
\Line(0,3)(60,63)
\Line(0,0)(60,60)
\Line(60,60)(120,0)
\Gluon(60,0)(60,50){4}{4}
\CCirc(60,60){10}{Black}{White}
 
\Gluon(30,30)(90,30){4}{4}
\end{picture}
&
\begin{picture}(90,50)(0,-5)
\SetScale{0.7}
\SetWidth{1}
\Line(0,3)(60,63)
\Line(0,0)(60,60)
\Line(60,60)(120,0)
\Gluon(60,0)(60,50){4}{4}
\GlueArc(35,35)(25,45,225){4}{4}
\CCirc(60,60){10}{Black}{White}
 
\end{picture}\\
A11 & A12 & A13 & A14\\
\end{tabular}
\end{center}
\begin{center}
\begin{tabular}{cccc} 
\begin{picture}(90,50)(0,-5)
\SetScale{0.7}
\SetWidth{1}
\Line(0,3)(60,63)
\Line(0,0)(60,60)
\Line(60,60)(120,0)
\Gluon(60,0)(60,15){4}{2}
\Gluon(60,35)(60,50){4}{2}
\CCirc(60,60){10}{Black}{White}

\CCirc(60,25){10}{Black}{Black} 
\end{picture} &
\begin{picture}(90,50)(0,-5)
\SetScale{0.7}
\SetWidth{1}
\Line(0,3)(60,63)
\Line(0,0)(60,60)
\Line(60,60)(120,0)
\Gluon(60,0)(60,50){4}{4}
\CCirc(60,60){10}{Black}{White}
 
\Gluon(60,30)(90,30){4}{3}
\end{picture}
&
\begin{picture}(90,50)(0,-5)
\SetScale{0.7}
\SetWidth{1}
\Line(0,3)(60,63)
\Line(0,0)(60,60)
\Line(60,60)(120,0)
\Gluon(60,0)(60,20){4}{2}
\GlueArc(60,35)(15,90,270){3}{4}
\GlueArc(60,35)(15,270,450){3}{4}
\CCirc(60,60){10}{Black}{White}
 
\end{picture} &
\begin{picture}(90,50)(0,-5)
\SetScale{0.7}
\SetWidth{1}
\Line(0,3)(60,63)
\Line(0,0)(60,60)
\Line(60,60)(120,0)
\Gluon(60,0)(60,50){4}{4}
\CCirc(60,60){10}{Black}{White}
 
\GlueArc(90,30)(20,-45,135){4}{4}
\end{picture}
\\
A22 & A23 & A24 & A33\\
\end{tabular}
\end{center}
\begin{center}
\begin{tabular}{cc}
\begin{picture}(90,50)(0,-5)
\SetScale{0.7}
\SetWidth{1}
\Line(0,3)(60,63)
\Line(0,0)(60,60)
\Line(60,60)(120,0)
\Gluon(60,0)(60,50){4}{4}
\GlueArc(85,35)(25,-45,135){4}{4}
\CCirc(60,60){10}{Black}{White}
 
\end{picture}
&
\begin{picture}(90,70)(0,-5)
\SetScale{0.7}
\SetWidth{1}
\Line(0,3)(60,63)
\Line(0,0)(60,60)
\Line(60,60)(120,0)
\Gluon(60,0)(60,50){4}{4}
\GlueArc(60,75)(15,-90,270){4}{8}
\CCirc(60,60){10}{Black}{White}
 
\end{picture}\\
A34 & A44
\end{tabular}
\end{center}
\caption{One-loop diagrams obtained from diagram A.}\label{fig:A}
\end{figure}

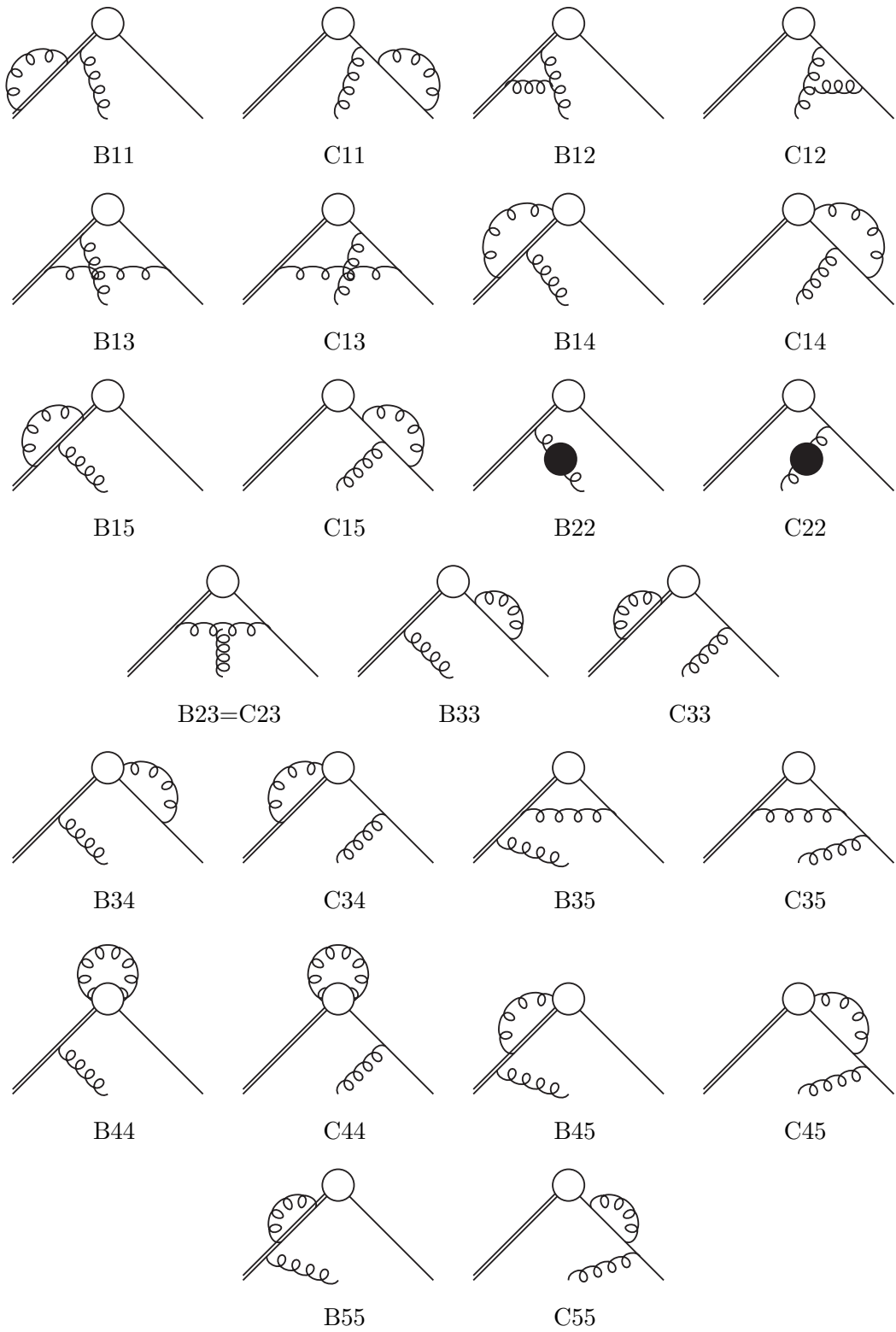
\begin{figure}[h]
\begin{center}
\begin{tabular}{cccc}
\begin{picture}(90,50)(0,-5)
\SetScale{0.7}
\SetWidth{1}
\Line(0,3)(60,63)
\Line(0,0)(60,60)
\Line(60,60)(120,0)
\Gluon(60,0)(45,45){4}{4}
\CCirc(60,60){10}{Black}{White}
 
\GlueArc(20,20)(20,45,225){4}{4}
\end{picture} & 
\begin{picture}(90,50)(0,-5)
\SetScale{0.7}
\SetWidth{1}
\Line(0,3)(60,63)
\Line(0,0)(60,60)
\Line(60,60)(120,0)
\Gluon(60,0)(75,45){4}{4}
\CCirc(60,60){10}{Black}{White}
 
\GlueArc(100,20)(20,-45,135){4}{4}
\end{picture} &
\begin{picture}(90,50)(0,-5)
\SetScale{0.7}
\SetWidth{1}
\Line(0,3)(60,63)
\Line(0,0)(60,60)
\Line(60,60)(120,0)
\Gluon(60,0)(45,45){4}{4}
\CCirc(60,60){10}{Black}{White}
 
\Gluon(20,20)(50,20){4}{3}
\end{picture} &
\begin{picture}(90,50)(0,-5)
\SetScale{0.7}
\SetWidth{1}
\Line(0,3)(60,63)
\Line(0,0)(60,60)
\Line(60,60)(120,0)
\Gluon(60,0)(75,45){4}{4}
\CCirc(60,60){10}{Black}{White}
 
\Gluon(100,20)(70,20){4}{3}
\end{picture}
\\
B11 & C11 & B12 & C12\\
\end{tabular}
\end{center}
\begin{center}
\begin{tabular}{cccc}
\begin{picture}(90,50)(0,-5)
\SetScale{0.7}
\SetWidth{1}
\Line(0,3)(60,63)
\Line(0,0)(60,60)
\Line(60,60)(120,0)
\Gluon(60,0)(45,45){4}{4}
\CCirc(60,60){10}{Black}{White}
 
\Gluon(20,20)(100,20){4}{4}
\end{picture} &
\begin{picture}(90,50)(0,-5)
\SetScale{0.7}
\SetWidth{1}
\Line(0,3)(60,63)
\Line(0,0)(60,60)
\Line(60,60)(120,0)
\Gluon(60,0)(75,45){4}{4}
\CCirc(60,60){10}{Black}{White}
 
\Gluon(20,20)(100,20){4}{4}
\end{picture} &
\begin{picture}(90,50)(0,-5)
\SetScale{0.7}
\SetWidth{1}
\Line(0,3)(60,63)
\Line(0,0)(60,60)
\Line(60,60)(120,0)
\Gluon(60,0)(35,35){4}{4}
\GlueArc(35,35)(25,45,225){4}{4}
\CCirc(60,60){10}{Black}{White}
 
\end{picture} & 
\begin{picture}(90,50)(0,-5)
\SetScale{0.7}
\SetWidth{1}
\Line(0,3)(60,63)
\Line(0,0)(60,60)
\Line(60,60)(120,0)
\Gluon(60,0)(85,35){4}{4}
\GlueArc(85,35)(25,-45,135){4}{4}
\CCirc(60,60){10}{Black}{White}
 
\end{picture}\\
B13 & C13 & B14 & C14\\
\end{tabular}
\end{center}
\begin{center}
\begin{tabular}{cccc}
\begin{picture}(90,50)(0,-5)
\SetScale{0.7}
\SetWidth{1}
\Line(0,3)(60,63)
\Line(0,0)(60,60)
\Line(60,60)(120,0)
\Gluon(60,0)(30,30){4}{4}
\CCirc(60,60){10}{Black}{White}
 
\GlueArc(30,30)(20,45,225){4}{4}
\end{picture} & 
\begin{picture}(90,50)(0,-5)
\SetScale{0.7}
\SetWidth{1}
\Line(0,3)(60,63)
\Line(0,0)(60,60)
\Line(60,60)(120,0)
\Gluon(60,0)(90,30){4}{4}
\CCirc(60,60){10}{Black}{White}
 
\GlueArc(90,30)(20,-45,135){4}{4}
\end{picture} &
\begin{picture}(90,50)(0,-5)
\SetScale{0.7}
\SetWidth{1}
\Line(0,3)(60,63)
\Line(0,0)(60,60)
\Line(60,60)(120,0)
\Gluon(70,0)(40,40){4}{4}
\CCirc(60,60){10}{Black}{White}

\CCirc(55,20){10}{Black}{Black}
\end{picture} &
\begin{picture}(90,50)(0,-5)
\SetScale{0.7}
\SetWidth{1}
\Line(0,3)(60,63)
\Line(0,0)(60,60)
\Line(60,60)(120,0)
\Gluon(50,0)(80,40){4}{4}
\CCirc(60,60){10}{Black}{White}
 
\CCirc(65,20){10}{Black}{Black}
\end{picture}
\\
B15 & C15 & B22 & C22\\
\end{tabular}
\end{center}
\begin{center}
\begin{tabular}{ccc}
\begin{picture}(90,50)(0,-5)
\SetScale{0.7}
\SetWidth{1}
\Line(0,3)(60,63)
\Line(0,0)(60,60)
\Line(60,60)(120,0)
\Gluon(60,0)(60,30){4}{4}
\CCirc(60,60){10}{Black}{White}
 
\Gluon(30,30)(90,30){4}{4}
\end{picture} &
\begin{picture}(90,50)(0,-5)
\SetScale{0.7}
\SetWidth{1}
\Line(0,3)(60,63)
\Line(0,0)(60,60)
\Line(60,60)(120,0)
\Gluon(60,0)(30,30){4}{4}
\CCirc(60,60){10}{Black}{White}
 
\GlueArc(85,35)(15,-45,135){4}{4}
\end{picture} &
\begin{picture}(90,50)(0,-5)
\SetScale{0.7}
\SetWidth{1}
\Line(0,3)(60,63)
\Line(0,0)(60,60)
\Line(60,60)(120,0)
\Gluon(60,0)(90,30){4}{4}
\CCirc(60,60){10}{Black}{White}
 
\GlueArc(35,35)(15,45,225){4}{4}
\end{picture}\\
B23=C23 & B33 & C33\\
\end{tabular}
\end{center}
\begin{center}
\begin{tabular}{cccc}
\begin{picture}(90,50)(0,-5)
\SetScale{0.7}
\SetWidth{1}
\Line(0,3)(60,63)
\Line(0,0)(60,60)
\Line(60,60)(120,0)
\Gluon(60,0)(30,30){4}{4}
\GlueArc(80,40)(20,-45,135){4}{4}
\CCirc(60,60){10}{Black}{White}
 
\end{picture} & 
\begin{picture}(90,50)(0,-5)
\SetScale{0.7}
\SetWidth{1}
\Line(0,3)(60,63)
\Line(0,0)(60,60)
\Line(60,60)(120,0)
\Gluon(60,0)(90,30){4}{4}
\GlueArc(40,40)(20,45,225){4}{4}
\CCirc(60,60){10}{Black}{White}
 
\end{picture} &
\begin{picture}(90,50)(0,-5)
\SetScale{0.7}
\SetWidth{1}
\Line(0,3)(60,63)
\Line(0,0)(60,60)
\Line(60,60)(120,0)
\Gluon(60,0)(15,15){4}{4}
\CCirc(60,60){10}{Black}{White}
 
\Gluon(30,30)(90,30){4}{4}
\end{picture} &
\begin{picture}(90,50)(0,-5)
\SetScale{0.7}
\SetWidth{1}
\Line(0,3)(60,63)
\Line(0,0)(60,60)
\Line(60,60)(120,0)
\Gluon(60,0)(105,15){4}{4}
\CCirc(60,60){10}{Black}{White}
 
\Gluon(30,30)(90,30){4}{4}
\end{picture}\\
B34 & C34 & B35 & C35\\
\end{tabular}
\end{center}
\begin{center}
\begin{tabular}{cccc}
\begin{picture}(90,70)(0,-5)
\SetScale{0.7}
\SetWidth{1}
\Line(0,3)(60,63)
\Line(0,0)(60,60)
\Line(60,60)(120,0)
\Gluon(60,0)(30,30){4}{4}
\GlueArc(60,75)(15,-90,270){4}{8}
\CCirc(60,60){10}{Black}{White}
 
\end{picture}&
\begin{picture}(90,70)(0,-5)
\SetScale{0.7}
\SetWidth{1}
\Line(0,3)(60,63)
\Line(0,0)(60,60)
\Line(60,60)(120,0)
\Gluon(60,0)(90,30){4}{4}
\GlueArc(60,75)(15,-90,270){4}{8}
\CCirc(60,60){10}{Black}{White}
 
\end{picture} &
\begin{picture}(90,50)(0,-5)
\SetScale{0.7}
\SetWidth{1}
\Line(0,3)(60,63)
\Line(0,0)(60,60)
\Line(60,60)(120,0)
\Gluon(60,0)(15,15){4}{4}
\GlueArc(40,40)(20,45,225){4}{4}
\CCirc(60,60){10}{Black}{White}
 
\end{picture} & 
\begin{picture}(90,50)(0,-5)
\SetScale{0.7}
\SetWidth{1}
\Line(0,3)(60,63)
\Line(0,0)(60,60)
\Line(60,60)(120,0)
\Gluon(60,0)(105,15){4}{4}
\GlueArc(80,40)(20,-45,135){4}{4}
\CCirc(60,60){10}{Black}{White}
 
\end{picture}\\
B44 & C44 & B45 & C45\\
\end{tabular}
\end{center}
\begin{center}
\begin{tabular}{cc}
\begin{picture}(90,50)(0,-5)
\SetScale{0.7}
\SetWidth{1}
\Line(0,3)(60,63)
\Line(0,0)(60,60)
\Line(60,60)(120,0)
\Gluon(60,0)(15,15){4}{4}
\CCirc(60,60){10}{Black}{White}
 
\GlueArc(35,35)(15,45,225){4}{4}
\end{picture} &
\begin{picture}(90,50)(0,-5)
\SetScale{0.7}
\SetWidth{1}
\Line(0,3)(60,63)
\Line(0,0)(60,60)
\Line(60,60)(120,0)
\Gluon(60,0)(105,15){4}{4}
\CCirc(60,60){10}{Black}{White}
\GlueArc(85,35)(15,-45,135){4}{4}
\end{picture}\\
B55 & C55\\
\end{tabular}
\end{center}
\caption{One-loop diagrams obtained from diagrams B and C.}\label{fig:BC}
\end{figure}

One can easily spell out the diagrams for $O_\pm$ at one loop by taking the above three
diagrams $ABC$, and adding a new gluon line in all possible ways (our naming scheme
reflects this idea). Diagram $A$ yields the diagrams in fig.~\ref{fig:A}, 
whereas the other diagrams coming from $B$ and $C$ are show in fig.~\ref{fig:BC}.
Let us notice the presence of redundant diagrams, namely $(B24)=(A12)$, $(C24)=(A23)$, $(B25)=(B12)$, $(C25)=(C12)$.

\subsection{Common contributions}

Since we are interested in the renormalisation properties of the
distribution amplitudes in the $\bar{\rm MS}$ scheme,
we quote here only the poles in $\varepsilon$ defined as
$d=4-2\varepsilon$ corresponding to ultraviolet divergences (we discuss our integration procedure on one example explicitly in App.~\ref{app:pole}). For the integrals going up to infinity, we keep the 
expression of the kernels before picking up the pole in $\varepsilon$, 
since the integration may give rise to double poles in 
the expression of $\gamma_{\pm,3}$, related to Sudakov logarithms.
We do not give the expressions corresponding to the wave-function renormalisation
of the external legs, i.e., $(11),(22),(33)$.

The following diagrams yield contributions of the same form for both distribution
amplitudes :
\begin{eqnarray}
(A12)&=&-\frac{\alpha_s}{4\pi}C_A 
        \left(\frac{1}{2}\right)g_s\epsilon_+  
     \ \bar{v}\dirac{n}_\pm\Gamma T^a u \  \frac{1}{\varepsilon}  
\int_{-q_+}^0 dl_+ \frac{1}{q_+ l_+ }
       \left[\delta\left(\omega-k_++l_+\right)
         -\delta\left(\omega-k_+\right)    
    \right]\nonumber\\
(A14)&=&-\frac{\alpha_s}{4\pi}(g_s\epsilon_+)
         \ \bar{v}\dirac{n}_\pm\Gamma T^a u \ 
         \frac{1}{\varepsilon} \int_0^\infty dl_+ 
            \left(\frac{l_+^2}{\mu^2}\right)^{-\varepsilon}\nonumber\\
 & \times&\left[
   \frac{2C_F}{l_+q_+}
      \left\{
    \left[\delta\left(\omega-k_+-q_+-l_+\right)
         -\delta\left(\omega-k_+-q_+\right)    
    \right]\right.
    +\left[\delta\left(\omega-k_+\right)
         -\delta\left(\omega-k_+-l_+\right)    
    \right]
   \right\}\nonumber\\
&& 
 - \frac{C_A}{q_+}
      \left\{\frac{1}{l_++q_+}
    \left[\delta\left(\omega-k_+-q_+-l_+\right)
         -\delta\left(\omega-k_+\right)    
    \right]
\left. 
-\frac{1}{l_+}
    \left[\delta\left(\omega-k_+-l_+\right)
         -\delta\left(\omega-k_+\right)    
    \right]
   \right\}
   \right]\nonumber\\
(A24) &=&-\frac{\alpha_s }{4\pi}C_A 
     \frac{(g_s\epsilon_+)}{2} \ \bar{v}\dirac{n}_\pm\Gamma T^a u \
     \frac{1}{\varepsilon}
      \int_{-q_+}^0 dl_+   \frac{2l_++q_+}{q_+(q_++l_+)} \nonumber\\
 &\times& \left[
   \frac{1}{l_+}
    \left[\delta\left(\omega-k_++l_+\right)
         -\delta\left(\omega-k_+\right)    
    \right]
+\frac{1}{q_+}
    \left[\delta\left(\omega-k_+-q_+\right)
         -\delta\left(\omega-k_+\right)    
    \right]\right]
   \nonumber\\
(A34) &=&\frac{\alpha_s}{4\pi}(g_s\epsilon_+)
         \ \bar{v}\dirac{n}_\pm\Gamma T^a u \ 
          \frac{1}{\varepsilon}
        \int^{k_+}_0 \frac{dl_+}{l_+} \frac{k_+-l_+}{k_+} \nonumber\\
 & \times&\left[
   \frac{2C_F}{q_+}
      \left\{
    \left[\delta\left(\omega-k_+-q_+\right)
         -\delta\left(\omega-k_+-q_++l_+\right)    
    \right]\right.
    +\left[\delta\left(\omega-k_++l_+\right)
         -\delta\left(\omega-k_+\right)    
    \right]
   \right\}\nonumber
\end{eqnarray}
\begin{eqnarray}
&&  - C_A \left\{\frac{1}{l_++q_+}
    \left[\delta\left(\omega-k_+-q_+\right)
         -\delta\left(\omega-k_++l_+\right)    
    \right]
\right.\nonumber\\
&& \qquad\qquad\qquad  \left. 
\left. 
-\frac{1}{q_+}
    \left[\delta\left(\omega-k_+-q_++l_+\right)
         -\delta\left(\omega-k_++l_+\right)    
    \right]
   \right\}
   \right]\nonumber\\
(A44)&=&0 \nonumber
\end{eqnarray}

For diagrams of $B$-type, we obtain:
\begin{eqnarray}
(B12) &=& 0\nonumber\\
(B13) &=& 0\nonumber\\
(B14) &=& 0\nonumber\\
(B15) &=& -\frac{\alpha_s}{4\pi} g_s \frac{1}{\varepsilon}\left(C_A-2C_F\right)\frac{v\cdot\epsilon}{v\cdot q}\delta\left(\omega - k_+\right)\, \bar{v} \dirac{n}_\pm\Gamma T^a u\nonumber\\
(B34) &=& B \otimes \left.Z_\pm\right|_{LO}\nonumber \\
(B35) &=& B \otimes \left.Z_\pm\right|_{LH}\nonumber \\
(B44) &=& 0\nonumber\\
(B45) &=& B \otimes \left.Z_\pm\right|_{HO}\nonumber \\
(B55) &=& B\times Z_h\nonumber
\end{eqnarray}

For diagrams of $C$-type, we obtain:
\begin{eqnarray}
(C12) &=&  \frac{\alpha_s}{4\pi}C_A \frac{3}{2}
           \frac{1}{(k+q)^2} 
         \ \bar{v}\dirac{\epsilon}(\dirac{k}+\dirac{q})\dirac{n}_\pm\Gamma T^a u \ 
          \frac{1}{\varepsilon}
    \delta\left(\omega-k_+-q_+\right)\nonumber\\ 
(C15) &=& \frac{\alpha_s}{4\pi}(C_F-C_A/2) 
           \frac{1}{(k+q)^2} 
         \ \bar{v}\dirac{\epsilon}(\dirac{k}+\dirac{q})\dirac{n}_\pm\Gamma T^a u \ 
          \frac{1}{\varepsilon} 
    \delta\left(\omega-k_+-q_+\right)\nonumber\\ 
(C34) &=& C \otimes \left.Z_\pm\right|_{HO}\nonumber \\
(C35) &=& C \otimes \left.Z_\pm\right|_{HL}\nonumber \\
(C44) &=& 0\nonumber \\ 
(C55) &=& C\times Z_q\nonumber
\end{eqnarray}

\subsection{$\phi_+$}
The remaining diagrams yield different contributions for $\phi_+$ and
$\phi_-$. For $\phi_+$ we have the following contributions:
\begin{eqnarray}
(A13_+)&=&0\nonumber\\
(A23_+)&=&\frac{\alpha_s}{4\pi}C_A\frac{(g_s\epsilon_+)}{2} \bar{v}\dirac{n}_+\Gamma T^a u\frac{1}{\varepsilon}
\left[\int_{-q_+}^0\frac{dl_+}{q_+}\left(k_+-q_+-2l_+\right)-2\int_0^{k_+}\frac{dl_+}{k_+}\left(k_+-l_+\right)\right]
\nonumber\\
&&
\qquad \times\frac{1}{(k_++q_+)(l_++q_+)}\left\{\delta(\omega-k_+-q_+)-\delta(\omega-k_++l_+)\right\}\nonumber\\
(B23_+)&=&0\nonumber\\
(C13_+)&=&0\nonumber\\
(C14_+)&=&-\frac{\alpha_s}{4\pi}(C_A-2C_F) g_s \frac{1}{\varepsilon} \bar{v}\dirac{n}_+\Gamma T^a u\epsilon_+
\left[\int_0^{k_+}\frac{dl_+}{k_+}+\int_{k_+}^{k_++q_+}\frac{dl_+}{q_+}\frac{k_++q_+-l_+}{l_+}\right]\nonumber\\
&&\qquad\times\frac{1}{k_++q_+}\left\{\delta(\omega-k_+-q_+)-\delta(\omega-k_+-q_++l_+)\right\}\nonumber\\
(C45_+)&=&C \otimes \left.Z_+\right|_{LO}\nonumber 
\end{eqnarray}
\newpage
\newpage
\subsection{$\phi_-$}

For $\phi_-$, the remaining diagrams yield the following contributions:
\begin{eqnarray}
(A13_-)&=&\frac{\alpha_s}{4\pi}\left(C_A-2C_F\right)g_s\epsilon_+
         \ \bar{v}\dirac{n}_-\Gamma T^a u \ 
             \frac{1}{\varepsilon} \nonumber\\
&&\qquad\times
           \int^{k_+}_0 dl_+ \frac{1}{q_+k_+}
       \left[\delta\left(\omega-k_+-q_++l_+\right)
         -\delta\left(\omega-k_++l_+\right)    
    \right]\nonumber\\
(A23_-) &=& \frac{\alpha_s }{4\pi}C_A 
     \frac{1}{2}g_s 
     \frac{1}{\varepsilon}
         \nonumber\\
&&\times \left\{\epsilon_+\ \bar{v}\dirac{n}_-\Gamma T^a u \ 
      \left[\int^{k_+}_0 \frac{dl_+}{k_+}\left(l_+-q_+-2k_+\left(1+ \frac{k_+-l_+}{k_++q_+}\right)\right)
       \right.\right.  \nonumber\\
&&\qquad\qquad\qquad\qquad\left.\left.  + \int_{-q_+}^0 \frac{dl_+}{q_+} \left(l_++k_+-2k_+ \frac{l_++q_+}{k_++q_+}\right)\right]\right.\nonumber\\
&&\qquad 
   + \epsilon_+ \bar{v}\dirac{q}_\perp \dirac{n}_+\dirac{n}_-\Gamma T^a
   u \ 
   \left[-\int^{k_+}_0 \frac{dl_+}{k_+} \frac{l_++q_+}{k_++q_+}
           + \int_{-q_+}^0 \frac{dl_+}{q_+}  \frac{(l_++q_+)(k_++2q_+)}{q_+(k_++q_+)}\right]
\nonumber\\
&&\qquad 
   +\frac{1}{2}  \bar{v}\dirac{\epsilon}_\perp \dirac{n}_+\dirac{n}_-\Gamma T^a
   u \ 
 \left.
   \left[\int^{k_+}_0 \frac{dl_+}{k_+} (q_++l_+-2k_+)
           - \int_{-q_+}^0 \frac{dl_+}{q_+}  (3q_++l_+)\right]\right\}\nonumber\\
&&
 \qquad \qquad\qquad \times  \frac{1}{(k_++q_+)(q_++l_+)}    
   \left[\delta\left(\omega-k_+-q_+\right)
         -\delta\left(\omega-k_++l_+\right)\right]\nonumber
\end{eqnarray}
\begin{eqnarray}
(B23_-) &=&-\frac{\alpha_s}{4\pi}\frac{C_A}{2}
    g_s
         \ \bar{v}[\epsilon_+\dirac{n}_-
                   -\frac{1}{2}\dirac{\epsilon}_\perp\dirac{n}_+\dirac{n}_-]\Gamma  T^a u \          
          \frac{1}{\varepsilon} \nonumber\\ 
&& \times\left[\int_{-q_+}^0 dl_+ \frac{1}{q_+(q_++k_+)}
        -\int^{k_+}_0 dl_+ \frac{1}{k_+(q_++k_+)}\right]
   \delta\left(\omega-k_++l_+\right)\nonumber\\ 
(C13_-) &=&\frac{\alpha_s}{4\pi}(C_A-2C_F)
       g_s\epsilon_+
         \ \bar{v}\dirac{n}_-\Gamma T^a u \ 
          \frac{1}{\varepsilon} \nonumber\\ 
&& \times\left[\int_{-q_+}^0 dl_+ \frac{1}{q_+(q_++k_+)}
        -\int^{k_+}_0 dl_+ \frac{1}{k_+(q_++k_+)}\right]
   \delta\left(\omega-q_+-l_+\right)\nonumber\\ 
(C14_-) &=& \frac{\alpha_s }{4\pi}(C_A-2C_F) 
     g_s 
     \frac{1}{\varepsilon}
         \nonumber\\
&&\times \Bigg\{
\epsilon_+\bar{v}\dirac{n}_-\Gamma T^a u\ 
\left[\int^{k_+}_0\frac{dl_+}{k_+}\left(l_+-k_+\frac{k_+-l_+}{k_++q_+}\right)-\int_{-q_+}^0\frac{dl_+}{q_+}k_+\frac{l_++q_+}{k_++q_+}\right]\nonumber\\
&&\qquad 
 +\frac{1}{2}  \epsilon_+ \bar{v}\dirac{q}_\perp\dirac{n}_+\dirac{n}_-\Gamma T^a
   u \ 
   \left[\int^{k_+}_0 \frac{dl_+}{k_+} \frac{k_+-l_+}{k_++q_+}
           + \int_{-q_+}^0 \frac{dl_+}{q_+} \left(\frac{l_+}{q_+}+ \frac{l_++q_+}{k_++q_+}\right)\right]
\nonumber\\
&&\qquad 
   +\frac{1}{2}  \bar{v}\dirac{\epsilon}_\perp\dirac{n}_+\dirac{n}_-\Gamma T^a
   u \ 
  \left. \left[\int^{k_+}_0 \frac{dl_+}{k_+} l_+
          - \int_{-q_+}^0 \frac{dl_+}{q_+} l_+\right]\right\}\nonumber\\
&&
 \qquad \qquad\qquad \times  \frac{1}{(k_+-l_+)(k_++q_+)}    
   \left[\delta\left(\omega-k_+-q_+\right)
         -\delta\left(\omega-q_+-l_+\right)    
    \right]\nonumber\\
(C45_-) &=& C \otimes \left.Z_-\right|_{LO}
- 2\frac{\alpha_s}{4\pi}C_F 
     g_s
         \ \bar{v}[\epsilon_+\dirac{n}_-
                   +\frac{1}{2}\dirac{\epsilon}_\perp\dirac{n}_+\dirac{n}_-]\Gamma  T^a u \ 
          \frac{1}{\varepsilon}
\nonumber\\ 
&& \qquad\qquad\qquad\qquad \times
\int_0^{k_++q_+} \frac{dl_+}{l_+} 
      \frac{k_++q_+-l_+}{(k_++q_+)^2}
\left[\delta\left(\omega-k_+-q_++l_+\right)
        -\delta\left(\omega-k_+-q_+\right)\right]\nonumber 
\end{eqnarray}

$(C45_-)$ contains an additional contribution compared to the 
two-parton case, because
the determination of $\left.Z_-\right|_{LO}$ relied on the fact that the
light-quark coming out of the vertex was on shell, which is not the case
for $(C45)$ in the three-parton case.

\section{One-loop mixing of $\phi_\pm$ with 3-parton distribution amplitudes}

\subsection{$\phi_+$}

Having calculated the divergent part of all possible diagrams 
the renormalisation matrix can be determined in a similar manner to ref.~\cite{Grozin:2005iz}. 
We write:
\begin{equation}
Z_\pm(\omega,\omega';\mu)=\delta(\omega-\omega')+\frac{\alpha_s(\mu)}{4\pi} z_\pm^{(1)}(\omega,\omega';\mu)
\qquad
Z_{\pm,3}(\omega,\omega',\xi';\mu)=\frac{\alpha_s(\mu)}{4\pi} z_{3\pm}^{(1)}(\omega,\omega',\xi';\mu),
\end{equation}
with $z_\pm^{(1)}$ being proportional to $C_F$.
One can schematically write for the matrix element of the bare operator up to
one loop:
\begin{eqnarray}\label{eq:rensum}
&&\langle 0\vert O_\pm(\omega)\vert H\rangle^{bare}
=Z_h^{1/2}Z_q^{1/2} Z_3^{1/2} Z_g [A+B+C]^{bare}\label{mat-el}\\
&&\qquad +[B34+B35+B45+C34+C35+C45]+[B12+B15+C12+C15+B55+C55]\nonumber\\ 
&&\qquad
   +[A12+A13+A14+A23+A24+A34+B23+C13+C14]\nonumber\\ 
&&= [A+B+C]^{ren}(\mu) \\
&&\qquad 
 +\frac{\alpha_s}{4\pi}\int d\omega' z^{(1)}_\pm(\omega,\omega';\mu)[A+B+C](\omega')
+\frac{\alpha_s}{4\pi}\int d\omega'd\xi' z_{3\pm}^{(1)\mu}(\omega,\omega',\xi';\mu) A_{3\mu}(\omega',\xi'),
\nonumber
\end{eqnarray}
where the renormalisation constants $Z_h$, $Z_q$, $Z_3$ and $Z_g$ 
come from the heavy-quark, light-quark
and gluon external legs and the coupling
constant respectively in the leading order contribution. 
Since the matrix element of the renormalized operator $O_\pm(\omega;\mu)$ 
must stay finite for
$\varepsilon\rightarrow0$ and since we know $z_\pm^{(1)}(\omega,\omega';\mu)$,
we can determine
$z_{3\pm}^{(1)}(\omega,\omega'\xi';\mu)$ from the poles of
the diagrams listed in (\ref{mat-el}). 

In the case of $B$ and $C$, the 
diagrams $(B34),(B35),(B45)$ and $(C34),(C35),(C45)$ together
with the fermion wave function renormalisation $Z_h^{1/2}$ 
and $Z_q^{1/2}$, given in eqs.~(\ref{eq:zpmwfrh})-(\ref{eq:zpmwfrl}), add up as indicated above to $B\otimes z_\pm^{(1)}$ and $C\otimes z_\pm^{(1)}$ respectively. 
The combination of the renormalisation constant for the coupling constant
and the gluon field tensor is:
\begin{equation}
Z_3^{1/2}Z_g = 1-\frac{\alpha_s C_A}{4\pi\epsilon}
\end{equation}
so that its contribution multiplied by $B$ and $C$ cancels the 
$C_A$-part of $(B12)+(B15)$ and $(C12)+(C15)$, whereas
 the $C_F$-part of the same
diagrams is cancelled by $(B55)$ and $(C55)$, as expected from general arguments on the
renormalisation of the quark-gluon vertex.

The remaining diagrams
in eq.~(\ref{eq:rensum}) must be added, and one has to subtract
$A\otimes z_\pm^{(1)}$ to extract the three-parton contribution, which
amounts to subtracting:
\begin{eqnarray}
A \otimes \left.Z_\pm\right|_{HO} &=&
 2\frac{\alpha_s C_F}{4\pi } \Gamma\left(\frac{1}{\varepsilon}\right)
    \times (-g_s) \frac{\epsilon_+}{q_+}
   \ \bar{v} \dirac{n}_\pm \Gamma  T^a u \  
    \int_0^\infty \frac{dl_+}{l_+} \left(\frac{l_+^2}{\mu^2}\right)^{-\varepsilon}\\
&& \qquad \times
       \left[\delta\left(\omega-k_+-q_+-l_+\right)
            -\delta\left(\omega-k_+-q_+\right)
          - \delta\left(\omega-k_+-l_+\right)
            +\delta\left(\omega-k_+\right)\right]\nonumber\\
A \otimes \left.Z_\pm\right|_{LO} &=&
2 \frac{\alpha_s C_F}{4\pi} \Gamma\left(\frac{1}{\varepsilon}\right)
    \times g_s \frac{\epsilon_+}{q_+}
   \ \bar{v} \dirac{n}_\pm \Gamma  T^a u \  
\\
&& \quad \times 
\left[\int_0^{q_+} \frac{dl_+}{l_+-q_+} \frac{l_++k_+}{q_++k_+}
              \left[\delta\left(\omega-k_+-l_+\right)-
                    \delta\left(\omega-k_+-q_+\right)
              \right]
\right.\nonumber\\
&& \qquad\qquad+ \int_{-k_+}^0 \frac{dl_+}{l_+-q_+} \frac{l_++k_+}{q_++k_+}
          \left[\delta\left(\omega-k_+-l_+\right)-
                    \delta\left(\omega-k_+-q_+\right)
              \right]
\nonumber\\
&& \qquad\qquad \left. -\int_{-k_+}^0 \frac{dl_+}{l_+} \frac{l_++k_+}{k_+}
          \left[\delta\left(\omega-k_+-l_+\right)-
                    \delta\left(\omega-k_+\right)
              \right]\right]\nonumber
\end{eqnarray}
\begin{eqnarray}
A \otimes \left.Z_-\right|_{HL} &=&
2 \frac{\alpha_s C_F}{4\pi} \frac{1}{\varepsilon} \times (-g_s) \frac{\epsilon_+}{q_+}
   \ \bar{v} \dirac{n}_- \Gamma  T^a u \  
\left[\int_0^{q_+} dl_+ \frac{1}{q_++k_+}
              \delta\left(\omega-k_+-l_+\right) \right.\\
&& \qquad \qquad \left. 
+\int_{-k_+}^0 dl_+ \frac{-q_+}{k_+(k_++q_+)} 
    \delta\left(\omega-k_+-l_+\right)\right] \nonumber\\
A \otimes \left. Z_+\right|_{HL}&=&0
\end{eqnarray}

Let us focus on $\phi_+$ in the remaining part of this section.
For the part proportional to $C_F$, the Sudakov-like contribution of $(A14)$
matches that of $Z_{HO}$ and more generally, 
an explicit computation shows that the diagrams in the fourth bracket
of  eq.~(\ref{eq:rensum})
add up exactly to the contribution from $z_+^{(1)}$ (which is proportional
to $C_F$). 

For the $C_A$ part, the contribution from $(A14)$ may seem
surprising at first glance, 
since it seems to involve another Sudakov-like integral, with a
double pole in $1/\epsilon$ generated by the integration of $l_+$ up to
infinity. But let us split the first integral in the 
$C_A$ term of $(A14)$ in two intervals, from 0 to $q_+$ and from $q_+$ to
$\infty$, perform 
a change of variable $l_+ \to l_+ + q_+$ and add the second integral, we
obtain:
\begin{eqnarray}
&&\frac{1}{\epsilon}\int_0^\infty dl_+
\left[\left(\frac{l_+}{\mu}\right)^{2\epsilon}-\left(\frac{l_++q_+}{\mu}\right)^{2\epsilon}\right]
\frac{1}{l_++q_+}
 \left[\delta\left(\omega-k_+-q_+-l_+\right)
         -\delta\left(\omega-k_+\right)    
    \right]\\
&&\qquad -\frac{1}{\epsilon}\int_0^{q_+} dl_+
  \left(\frac{l_+}{\mu}\right)^{2\epsilon}\frac{1}{l_+}
    \left[\delta\left(\omega-k_+-l_+\right)
         -\delta\left(\omega-k_+\right)    
    \right]\nonumber
\end{eqnarray}
One can see that the first integral yields no pole in $1/\epsilon$, and
the second integral provides a simple pole. Summing up all the contributions
proportional to $C_A$ (including that 
of the renormalisation constants from the  gluon field tensor and the strong
coupling constant), one observes that they cancel exactly.

In summary, the determination of the renormalisation properties of $\phi_+$ at
one loop with a three-parton external state
yields a $C_F$ term equal to the self-mixing obtained 
from the consideration of two-parton external state, and no $C_A$ term. This
shows that $\phi_+$ mixes only with itself, and  not with 
three-parton distribution amplitudes, up to one loop. 

The fact that $\phi_+$ occurs in most of the factorisation analyses for B-meson decays suggests that it holds a special status with respect to other $B$-meson distribution amplitudes, which is somehow confirmed by our finding of an absence of mixing. For light mesons, conformal symmetry would naturally explain the absence of mixing between distribution with different parton numbers and thus different twists~\cite{Braun:2003rp}. In the heavy-quark case, conformal symmetry cannot be invoked anymore~\cite{Braun:2003wx}, but our result may be the hint of another symmetry singling out $\phi_+$ with respect to other distribution amplitudes and explaining that $\gamma_{+,3}=0$.

\subsection{$\phi_-$}

We consider now $\phi_-$. One can follow the same argument as before, with a
similar pattern of cancellation for the diagrams yielding the same results
for $\phi_+$ and $\phi_-$. In particular,
one recovers easily the contribution proportional to $Z_-$, i.e. the
contribution from self-mixing derived from the two-particle case.

But the diagrams do not cancel completely, and there remains a genuine three-particle contribution:
\begin{eqnarray}\label{eq:diagcont}
\langle 0\vert O_-(\omega)\vert H\rangle^{bare}
&=& [A+B+C]^{ren}(\mu) 
 +\frac{\alpha_s}{4\pi}\int d\omega' z^{(1)}_\pm(\omega,\omega';\mu)[A+B+C](\omega')\nonumber\\
&& +\frac{1}{2}\frac{\alpha_s}{4\pi}g_s\frac{1}{\varepsilon}\left[q_+\bar{v}(k)[\dirac{\epsilon}_\perp(q)\dirac{n}_+\dirac{n}_-\Gamma T^a]u(p)-\bar{v}(k)[\dirac{q}_\perp\dirac{n}_+\dirac{n}_-\Gamma T^a]u(p)\epsilon_+(q)\right]\nonumber\\
&&\quad\times\left\{(C_A-2C_F)\left[\frac{1}{q_+^2}\int_{k_+}^{k_++q_+}dl_+\left(\frac{1}{l_+}-\frac{1}{k_+}\right)+\frac{1}{(k_++q_+)^2}\int_0^{k_++q_+}\frac{dl_+}{k_+}\right]\right.\nonumber\\
&&\qquad\qquad\qquad-\left.C_A\frac{1}{k_+}\left[\frac{1}{(k_++q_+)^2}\int_0^{k_++q_+}dl_+-\frac{1}{q_+^2}\int_0^{q_+}dl_+\right]\right\}\nonumber\\
&&\qquad\qquad\qquad\qquad\qquad\times\left\{\delta(\omega-k_+-q_++l_+)-\delta(\omega-k_+-q_+)\right\},
\end{eqnarray}
from which $z_{-,3}^{(1)\mu}$ can be extracted. 
If we separate the Dirac structure 
\begin{equation}
z_{-,3}^{(1)\mu}(\omega,\omega',\xi';\mu)=z_{-,3}^{(1)}(\omega,\omega',\xi';\mu)\gamma_\perp^\mu\dirac{n}_+\dirac{n}_-
\end{equation}
we obtain 
\begin{eqnarray}
z_{-,3}^{(1)}(\omega,\omega',\xi';\mu)&=&z_{-,3}^{(1)}(\omega,\omega',\xi')\nonumber\\
&=&-\frac{i}{2\varepsilon}\left[\frac{\Theta(\omega)}{\omega'}\left\{(C_A-2C_F)\left[\frac{1}{\xi'^2}\frac{\omega-\xi'}{\omega'+\xi'-\omega}\Theta(\xi'-\omega)+\frac{\Theta(\omega'+\xi'-\omega)}{(\omega'+\xi')^2}\right]\right.\right.\nonumber\\
&&\qquad -\left.\left.C_A\left[\frac{\Theta(\omega'+\xi'-\omega)}{(\omega'+\xi')^2}-\frac{1}{\xi'^2}\left(\Theta(\omega-\omega')-\Theta(\omega-\omega'-\xi')\right)\right]\right\}\right]_+.
\end{eqnarray}
We defined the +-distribution as:
\begin{equation}
\Big[f(\omega,\omega',\xi')\Big]_+ =f(\omega,\omega',\xi')-\delta(\omega-\omega'-\xi')\int
d\omega f(\omega,\omega',\xi'')
\end{equation}
Inserting $\gamma_\perp^\mu \dirac{n}_+\dirac{n}_-$ into the definition of the
three-particle distribution amplitudes one obtains the following expression
\begin{equation}
O_3(\omega',\xi')\,=\, 2(2-D)\left(\Psi_A(\omega',\xi')-\Psi_V(\omega',\xi')\right),
\end{equation}
which is exactly the combination arising in the constraint derived from the
equation of motion of the light quark in ref.~\cite{Kawamura:2001jm}.
At order $\alpha_s$, the other three-particle distribution amplitudes do not
mix with $\phi_-$.

One may use that to order $\alpha_s$ the following relations hold~\cite{Grozin:2005iz}:
\begin{eqnarray}
\frac{\partial O_-(\omega;\mu)}{\partial \log\mu}&=&
  -\int d\omega'\frac{\partial Z_-(\omega,\omega';\mu)}{\partial \log\mu} O_-(\omega';\mu)
-\int d\omega'd\xi'\frac{\partial Z_{-,3} (\omega,\omega',\xi';\mu)}{\partial\log\mu} O_3(\omega',\xi';\mu),\\
\frac{\partial \phi_-(\omega;\mu)}{\partial \log\mu}&=&-\frac{\alpha_s(\mu)}{4\pi}\left(\int d\omega \gamma^{(1)}_-(\omega,\omega';\mu)\phi_-(\omega';\mu)
+\int d\omega'd\xi' \gamma_{-,3}^{(1)}(\omega,\omega',\xi';\mu) [\Psi_A-\Psi_V](\omega',\xi';\mu)\right)\nonumber
\end{eqnarray}
and
\begin{equation}
\frac{\partial Z_{-,3}(\omega,\omega',\xi';\mu)}{\partial \log \mu}\,=\,-2\varepsilon\frac{\alpha_s(\mu)}{4\pi } z_{-,3}^{(1)}(\omega,\omega',\xi').
\end{equation}
taking into account that $Z_{-,3}$ starts only at $O(\alpha_s)$.
This leads to the anomalous dimension $\gamma_{-,3}^{(1)}$:
\begin{eqnarray}
\gamma_{-,3}^{(1)}(\omega,\omega',\xi';\mu)&=&\gamma_{-,3}^{(1)}(\omega,\omega',\xi')\nonumber\\
&=&4\left[\frac{\Theta(\omega)}{\omega'}\left\{(C_A-2C_F)\left[\frac{1}{\xi'^2}\frac{\omega-\xi'}{\omega'+\xi'-\omega}\Theta(\xi'-\omega)+\frac{\Theta(\omega'+\xi'-\omega)}{(\omega'+\xi')^2}\right]\right.\right.\nonumber\\
&&\qquad-\left.\left.C_A\left[\frac{\Theta(\omega'+\xi'-\omega)}{(\omega'+\xi')^2}-\frac{1}{\xi'^2}\left(\Theta(\omega-\omega')-\Theta(\omega-\omega'-\xi')\right)\right]\right\}\right]_+
\end{eqnarray}
corresponding to the one-loop mixing between $\phi_-$ and $\Psi_A-\Psi_V$.

\section{Calculation in a general covariant gauge}

We have computed the mixing between gauge-invariant operators, and we could in
principle have chosen any gauge to perform our determination of the renormalisation
properties of the latter. We can check the validity of our previous
computations by computing the nontrivial diagrams considered previously in a
general covariant gauge, where we replace the Feynman-gauge gluon-propagator by
\begin{equation}
d_{\mu\nu}^{ab}(k)=-i\delta^{ab}\left[g_{\mu\nu}-(1-\alpha)\frac{k_\mu\,k_\nu}{k^2}\right]
\end{equation}
Our result should be gauge invariant, so that the parts proportional to
$(1-\alpha)$ should cancel.

\subsection{Two-parton external state and $Z_\pm$}

First, we can repeat the computation of refs.~\cite{Lange:2003ff,Bell:2008er}, recalled in
sec.~\ref{sec:twopart}, for the case of a two-parton external state. The gauge
dependent part is:
 \begin{eqnarray}
M_{HO\pm}&\to&-ig_s^2 C_F (1-\alpha)\int \frac{d^4l}{(2\pi)^4} \bar{v}\dirac{n}_\pm\Gamma T^a u\frac{1}{l^4}\left\{\delta(\omega-k_+-l_+)-\delta(\omega-k_+)\right\},\\
M_{LO\pm}&\to&-ig_s^2 C_F (1-\alpha)\int \frac{d^4l}{(2\pi)^4} \bar{v}\dirac{n}_\pm\Gamma T^a u\frac{1}{l^4}\left\{\delta(\omega-k_++l_+)-\delta(\omega-k_+)\right\},\\
M_{HL\pm}&\to&ig_s^2 C_F (1-\alpha)\int \frac{d^4l}{(2\pi)^4} \bar{v}\dirac{n}_\pm\Gamma T^a u\frac{1}{l^4}\delta(\omega-k_++l_+).
 \end{eqnarray}
These integrals should be equipped with an infrared regulator (for instance, a
gluon mass $m_g$) in order to ensure that we keep only the ultraviolet divergences of interest 
here when we pick up the poles in $\varepsilon$ (otherwise, dimensional regularisation would treat
both ultraviolet and infrared divergences of the integral as poles in $\varepsilon$).

In ref.~\cite{Yan:1973qg}, such integrals were considered with a particular focus
on the integration over the different light-cone components $d^4l\rightarrow
\frac{1}{2} dl_+ dl_- d^2l_\perp$. Let us suppose that we want to integrate
over $l_-$. There is a single pole, at
$l_-=(m_g^2-i\epsilon+\vec{l}_\perp^2)/l_+$, that we can always avoid by
choosing the contour from above for $l_+>0$ and from below for $l_+<0$. It
seems to indicate that such integral should be 0, which is incorrect. As
proposed in ref.~\cite{Yan:1973qg}, a proper regularisation leads to the
conclusion that an integration over the minus
(plus) component results in a delta-distribution $\delta(l_+)$
($\delta(l_-)$), and one gets the equality:
\begin{equation}
\int \frac{d^4l}{(2\pi)^4} \frac{1}{l^4}\delta(\omega-k_+\pm l_+)\:=\:\delta(\omega-k_+)\int \frac{d^4l}{(2\pi)^4} \frac{1}{l^4}
\label{Yan-eq}
\end{equation}

We see that $M_{HO}$ and $M_{LO}$ vanish, whereas $M_{HL}$ cancels the gauge-dependent part of the wave-function renormalisation for the heavy and the
light quarks:
\begin{equation}
Z_q=1+C_F\frac{\alpha_s}{4\pi}\frac{1}{\varepsilon}[-1+(1-\alpha)] \qquad Z_h=1+C_F\frac{\alpha_s}{4\pi}\frac{1}{\varepsilon}[2+(1-\alpha)]
\end{equation}
Therefore, the gauge-dependent parts of the different contributions cancel and
we have checked that the expression of $Z_\pm$ is indeed gauge independent.

\subsection{Three-parton external state and $Z_{-3}$}

The issue becomes a little more involved if a three-particle state is
considered. The complete formulae can be found in App.~\ref{app:gencov}, but we can
outline the pattern of cancellation for the gauge-dependent part among the
various diagrams. 

We can identify the different gauge-dependent contributions
in eq.~(\ref{eq:diagcont}). The diagrams $(A44)$, $(B44)$ and $(C44)$, which vanish
trivially in the Feynman gauge,
have to be taken into account, but their contributions can be shown to vanish
through eq.~(\ref{Yan-eq}). This is also the case for the diagrams
$(A14)$, $(A24)$ and $(A34)$. $(B13)$ and $(B14)$ remain finite as in the
Feynman gauge.  

In analogy with the two-particle case, the diagrams
$(B34)$, $(B35)$, $(B45)$ and $(C34)$, $(C35)$, $(C45)$ cancel with
the gauge-dependent part of $Z_q^{1/2} Z_h^{1/2}$ multiplied by $B$ and $C$
respectively. One has to pay attention to 
 $(C35)$ and $(C45)$ that give additional contributions canceling each other.
$(B12)$, $(B15)$, $(B55)$ and $(C12)$, $(C15)$, $(C55)$ cancel against the
gauge-dependent part of $Z_3^{1/2}Z_g$ multiplied by $B$ and $C$ respectively.

Finally, one is left with $(A12)$, $(A13)$, $(A23)$, $(B23)$,
$(C13)$, $(C14)$ and $Z_h^{1/2}Z_2^{1/2}Z_g Z_3^{1/2}\times A$. The sum of $(C13)$ and $(C14)$ is finite,
and once eq.~(\ref{Yan-eq}) is applied, only the following expressions remain: 
\begin{eqnarray}
(A12)&=&-ig_s^3\frac{C_A}{4}(1-\alpha)\int\frac{d^4l}{(2\pi)^4}\frac{1}{l^4}
\frac{1}{q_+}\left\{\delta(\omega-k_+-q_+)-\delta(\omega-k_+)\right\}
   \bar{v}\dirac{n}_\pm\Gamma T^a u\epsilon_+,\\
(A13)&=&ig_s^3\left(\frac{C_A}{2}-C_F\right)(1-\alpha)\int\frac{d^4l}{(2\pi)^4}\frac{1}{l^4}
\frac{1}{q_+}\left\{\delta(\omega-k_+-q_+)-\delta(\omega-k_+)\right\}
\bar{v}\dirac{n}_\pm\Gamma T^a u\epsilon_+,\\
(A23)&=&-ig_s^3\frac{C_A}{4}(1-\alpha)\int\frac{d^4l}{(2\pi)^4}\frac{1}{l^4}\frac{1}{q_+}\left\{\delta(\omega-k_+-q_+)-\delta(\omega-k_+)\right\}
\bar{v}\dirac{n}_\pm\Gamma T^a u\epsilon_+,\\
(B23)&=&-ig_s^3\frac{C_A}{4}(1-\alpha)\int\frac{d^4l}{(2\pi)^4}\frac{1}{l^4}\frac{1}{q_+}\left\{\delta(\omega-k_+-q_+)-\delta(\omega-k_+)\right\}
\bar{v}\dirac{n}_\pm\Gamma T^a u\epsilon_+.
\end{eqnarray}
Picking up the ultraviolet divergences from the integrals, we obtain finally for the sum:
\begin{equation}
\frac{\alpha_s}{4\pi}\left(\frac{C_A}{4} +C_F\right)(1-\alpha)\frac{1}{\varepsilon}\bar{v}\dirac{n}_\pm\Gamma T^a u\epsilon_+\frac{1}{q_+}\left\{\delta(\omega-k_+-q_+)-\delta(\omega-k_+)\right\}
\end{equation}
which cancels exactly the $(1-\alpha)$-dependent part of the combination
\begin{equation}
Z_h^{1/2}Z_2^{1/2}Z_g Z_3^{1/2}\times A\:=\:\left(1+\frac{\alpha_s}{4\pi}\frac{1}{\varepsilon}
 \left(\frac{C_F}{2}[1+2(1-\alpha)]-\frac{C_A}{4}[4-(1-\alpha)]\right)\right)\times A,
\end{equation}
The calculation for $\phi_+(\omega)$ is simpler, since the replacement of
$\dirac{n}_-$ by $\dirac{n}_+$ implies the absence of
contributions proportional to $\bar{v} \dirac{\epsilon}
\dirac{n}_+\dirac{n}_-\Gamma T^a u$. Therefore the additional terms of $(C35)$ and
$(C45)$ vanish as well as the gauge-dependent contributions of the diagrams
$(C13)$ and $(C14)$. Following the same lines in both cases, one can conclude that there is no
gauge-dependence for the renormalisation of $\phi_+$ and $\phi_-$.

\section{Conclusion}

In this paper, we have studied the mixing of both two-particle distribution amplitudes $\phi_+$ and $\phi_-$ with
three-particle ones up to one-loop. Using the fact that RGE is a short-distance property of the operator,
we used matrix elements of the operators with a quark-antiquark-gluon external state. Determining the ultraviolet divergences
of the corresponding diagrams allowed us to recover the known one-loop self-mixing of $\phi_+$ and $\phi_-$, but also to determine
the role of three-parton distribution amplitudes. We have established that $\phi_+$ mixes only with itself, whereas $\phi_-$ does mix
with $[\Psi_A-\Psi_V]$, and we have provided the corresponding anomalous dimension. Through the use of a general covariant gauge, we have checked that our results were indeed gauge invariant, providing further support to our expressions.

We can relate our results to other comments on the $B$-meson distribution amplitudes in the literature. For instance,
the fact that $\phi_+$ does not mix with three-parton distribution amplitudes was already presented in ref.~\cite{Kawamura:2008vq}.
In this article, a computation similar to ours was sketched in the case of $\phi_+$, with the conclusion (presented in eq.~(2) of this reference) that the only ultraviolet one-loop divergence for $O_+$ is proportional to itself, whereas contributions proportional to higher-dimension operators have only infrared divergences.

As mentioned in the introduction, the presence of $\delta(\omega-\omega')\log(\mu/\omega)$ in the renormalization-matrices provides a radiative tail to $\phi_\pm$ falling off like $(\log \omega)/\omega$ for large $\omega$. It requires
one to consider either negative moments of the distribution amplitudes $\phi_-,\,\phi_+$, or 
positive moments with an ultraviolet cut-off\cite{Lange:2003ff,Lee:2005gza, Grozin:2005iz, Bell:2008er,Kawamura:2008vq}:
\begin{equation}
\langle \omega^N\rangle_\pm(\mu)\,=\,\int_0^{\Lambda_{UV}} d\omega\,\omega^N\,\phi_\pm(\omega;\mu)
\end{equation}
On the contrary, it is interesting to notice the limit
\begin{equation}
\lim_{\Lambda_{UV}\to \infty}\int_0^{\Lambda_{UV}} d\omega\, \omega^N\, z_{-,3}^{(1)}(\omega,\omega',\xi')=0\qquad N=0,\,1
\end{equation}
This is relevant for the calculation of the three-particle contributions to the moments:
\begin{eqnarray}
\int_0^{\Lambda_{UV}}d\omega\, \omega^N\,\phi_-(\omega;\mu) &=& 1+\frac{\alpha_s}{4\pi}\left(\int d\omega'\,\phi_-(\omega')\,\int_0^{\Lambda_{UV}}d\omega\, \omega^N\,z^{(1)}_-(\omega,\omega';\mu)\right.\\
&&-\left.\int d\omega'd\xi'(2-D)[\Psi_A-\Psi_V](\omega',\xi')\int_0^{\Lambda_{UV}}d\omega\,\omega^N\,z^{(1)}_{-,3}(\omega,\omega',\xi')\right)\nonumber
\end{eqnarray}
Therefore, there is no contribution to the two lowest moments of
$\phi_-$ from three-particle distribution amplitudes, which confirms the
statement made after eq.~(62) in ref.~\cite{Bell:2008er}. We have explicitly checked that this property of $z_{-,3}^{(1)}$ does not hold for higher positive moments $(N\geq 2)$.

In
ref.~\cite{Kawamura:2001jm} (see also refs.~\cite{Geyer:2005fb,Geyer:2007yh,Huang:2005kk}) were derived two different relations between $\phi_+$ and
$\phi_-$ on one hand and
the four three parton-distribution amplitudes $\Psi_V$, $\Psi_A$, $X_A$ and
$Y_A$ on the other hand:
\begin{eqnarray}\label{eq:i}
\omega \phi_-'(\omega)+\phi_+(\omega)&=&I(\omega)\\
(\omega-2\bar\Lambda)\phi_+(\omega)+\omega\phi_-(\omega)&=&J(\omega)\label{eq:j}
\end{eqnarray} 
$I$ $(J)$ is an integro-differential expression involving
$\Psi_A-\Psi_V$ ($\Psi_A+X_A$ and $\Psi_V$).
The first relation comes from the equation of motion for the light quark and the
latter one from the heavy quark (as suggested by the presence of the 
HQET parameter $\bar{\Lambda}=M_B-m_b$). The use of the equation of motion of the heavy quark
was criticised in ref.~\cite{Braun:2003wx,Bell:2008er}, because this is linked to the heavy-quark limit
which does not commute with the light-cone limit. Moreover, this equation can be derived only if one leaves the light-cone limit, which is not needed for the first relation. One can notice that the shapes of
the distribution amplitudes have been derived in the Wandura-Wilczek
approximation, where three-parton distribution amplitudes are
neglected ($I=J=0$), leading to rather unphysical shapes for the distribution amplitudes. If we assume that the three-particle distribution amplitudes mix separately into themselves~\cite{DescotesGenon:2009mt,DescotesOffen}, 
our work shows that the renormalisation-scale dependence of eq.~(\ref{eq:i}), derived from the light-quark equation of motion, is satisfying: by applying $d/d\log\mu$ to the equation of motion, both $\phi_-$ and $\Psi_A-\Psi_V$
yield a term proportional to $\Psi_A-\Psi_V$. The contributions proportional to the two-particle distribution amplitudes
 were shown to cancel in the Wandzura-Wilczek approximation in App.~D of
 ref.~\cite{Bell:2008er} (eq.~(\ref{eq:i}) was also shown to hold in a specific non-relativistic model
 for $\phi_+,\phi_-,\Psi_A,\Psi_V$ in App.~C of the same reference).
Eq.~(\ref{eq:j}) does not seem to have such a satisfactory renormalisation-scale
dependence, which would add to the various criticisms raised against this equation (see ref.~\cite{DescotesGenon:2009mt} for further discussion of this issue).

More generally, the influence of three-particle distribution-amplitudes on $\phi_-(\omega;\mu)$ requires one to model them. However, the only available models~\cite{Khodjamirian:2006st} assume $\Psi_A(\omega,\xi)=\Psi_V(\omega,\xi)$ and they yield no contribution to the evolution of $\phi_-(\omega;\mu)$. For practical calculations as well as for further model-building 
of distribution amplitudes beyond  $\phi_+$,
one needs the evolution kernel of the three-particle distribution amplitudes, which will be the subject of a future work~\cite{DescotesOffen}.

\section*{Acknowledgments}

We thank Thorsten Feldmann for useful discussions. Work supported in part by EU Contract No. 
  MRTN-CT-2006-035482, \lq\lq FLAVIAnet'' and
 by the ANR contract  \lq\lq DIAM'' ANR-07-JCJC-0031.
\begin{appendix}

\section{Extraction of poles related to UV divergences}\label{app:pole}

In this paper, we compute various integrals in dimensional regularisation to extract the $\varepsilon$-poles
related to UV divergences. The textbook procedure consists in a covariant anlaysis, where all space directions are treated on the same footing. Since we use light-cone coordinates with a privileged direction for the definition and discussion of the distribution amplitudes, we use a slightly less usual method which we apply on the illustrative integral:
\begin{equation}
I=\int \frac{d^4l}{(2\pi)^4} f(l_+) \frac{1}{l^2} \frac{1}{(l-k)^2}
\end{equation}
where $f$ is an arbitrary function of $l_+$ alone, corresponding to a gluon line (of momentum $l$) attached to a light-quark line (of incoming momentum $k$). Such an integral is needed already to compute the mixing of $\phi_\pm$ into themselves ($Z_{LO}$). 
We want to perform the integrals over $l_-$ and 
$\vec{l}_\perp$ in $4-2\varepsilon$ dimensions and isolate the poles in $\varepsilon$ related to UV 
divergences. We therefore introduce a small mass $m$ for the light quark to regularise (soft) IR divergences that are of not interest for the determination of the RG properties of the distribution amplitudes.

We perform first the integral over $l_-$ by identifying the poles in the complex $l_-$ plane :
\begin{equation}
I=\int \frac{dl_+ dl_- d^2\vec{l}_\perp}{2(2\pi)^4} f(l_+) \frac{1}{l^2+i0^+} \frac{1}{l^2-2k\cdot l+i0^+}
\end{equation}
which are $l_-=(\vec{l}_\perp^2-i0^+)/l_+$ and $l_-=k_-+((\vec{l}_\perp-\vec{k}_\perp))^2-i0^+)/(l_+-k_+)$, whose positions with respect to the real axis depend on the value of $l_+$. If $l_+$ is negative or larger than $k_+$, the two poles sit on the same side, and thus the contour integral yields 0. If $0<l_+<k_+$, the two poles are on different sides and one gets a non-vanishing contribution, for instance by closing the contour in the lower half-plane and thus picking up the first pole (associated with the $l^2$-denominator):
\begin{equation}
I=\frac{i}{4\pi}\int_0^{k_+} dl_+ \int \frac{d^2 \vec{l}_\perp}{(2\pi)^2}
    \frac{f(l_+)}{k_- l_+^2+k_+ \vec{l}_\perp^2- 2\vec{k}_\perp\cdot \vec{l}_\perp l_+}
\end{equation}
Then one can perform the integral over the $2-2\varepsilon$ transverse dimensions, which yields
the result:
\begin{equation}
I=\Gamma(\varepsilon)\frac{i}{(4\pi)^{2-\varepsilon}}\int_0^{k_+}\frac{dl_+}{k_+} f(l_+)
  \left[\frac{m^2}{\mu^2}\frac{l_+^2}{k_+^2}\right]^{-\varepsilon}
\end{equation}
This expression yields a single pole in $\varepsilon$ corresponding to the UV-divergent part of the integral, which enters $Z_{LO}$. The same procedure is applied to all the diagrams, with sometimes more involved integrals (up to four propagators), leading to the results quoted in the present article.

\section{Gauge dependence of the diagrams in a general covariant gauge} \label{app:gencov}

In this appendix, we collect the integrals from the different diagrams that are proportional to the
gauge parameter $(1-\alpha)$. For the well known cases (B12), (B15), (B55), (C12), (C15) and (C55), corresponding to the vertex renormalisation, the integrals have already been carried out.
\begin{eqnarray}
(A12)&=&-ig_s^3(1-\alpha)\frac{C_A}{2}\bar{v}\dirac{n}_\pm\Gamma T^a u \epsilon_+\nonumber\\
&&\times\int \frac{d^4l}{(2\pi)^4}\frac{1}{l_++q_+}\left\{\delta(\omega-k_+-q_+-l_+)-\delta(\omega-k_+)\right\}
\nonumber\\
&&\times
\left[\left(1-\frac{l_++q_+}{2q_+}\right)\frac{1}{l^4}+\frac{l_++q_+}{2q_+}\frac{1}{(l+q)^4}\right]\nonumber\\
(A13)&=&\frac{i}{2}g_s^3(1-\alpha)(C_A-2C_F)\bar{v}\dirac{n}_\pm\Gamma T^a u \epsilon_+
\nonumber\\
&&\times
\int\frac{d^4l}{(2\pi)^4} \frac{1}{l^4}\frac{1}{q_+}\left\{\delta(\omega-k_+-q_++l_+)-\delta(\omega-k_++l_+)\right\}\nonumber\\
(A14)&=&ig_s^3(1-\alpha)\bar{v}\dirac{n}_\pm\Gamma T^a u \epsilon_+\int \frac{d^4l}{(2\pi)^4}\frac{1}{l^4}\nonumber\\
&&\times \left\{C_F\frac{1}{q_+}\left(\delta(\omega-k_+-q_+-l_+)-\delta(\omega-k_+-q_+)-\right.
\right. \nonumber\\
&&-\left.
\delta(\omega-k_+-l_+)+\delta(\omega-k_+)\right)\nonumber\\
&&-\frac{C_A}{2}\frac{1}{q_+}\left(\frac{l_+}{l_++q_+}\left\{\delta(\omega-k_+-q_+-l_+)-\delta(\omega-k_+)\right\}
\right. \nonumber\\
&& \left.
-\delta(\omega-k_+-l_+)+\delta(\omega-k_+)\right)\Bigg\}\nonumber\\
(A23)&=&-ig_s^3(1-\alpha)\frac{C_A}{2}\dint{4}{l} \frac{1}{l_++q_+}\left\{\delta(\omega-k_+-q_+)-\delta(\omega-k_++l_+)\right\}\nonumber\\
&&\times\left[\left(\frac{1}{l^4}\left(1-\frac{1}{2}\frac{l_++q_+}{q_+}\right)+\frac{1}{2}\frac{l_++q_+}{q_+}\frac{1}{(l+q)^4}\right)\bar{v}\dirac{n}_\pm\Gamma T^a u\epsilon_+\right.\nonumber\\
&&\left.+\frac{1}{2}\frac{l_++q_+}{k_++q_+}\left(\frac{1}{(k-l)^2(l+q)^2}-\frac{1}{(l+q)^4}\right)\bar{v}\dirac{\epsilon}\dirac{n}_+\dirac{n}_\pm\Gamma T^a u\right]\nonumber\\
(A24)&=&+ig_s^3(1-\alpha)\frac{C_A}{2}\bar{v}\dirac{n}_\pm\Gamma T^a u\epsilon_+
\nonumber\\
&&\times
\int \frac{d^4l}{(2\pi)^4}\left[\frac{1}{(l+q)^4}\left(1+\frac{l_+}{2q_+}\right)+\frac{l_+}{l^4}\left(\frac{1}{l_++q_+}-\frac{1}{2q_+}\right)\right]\nonumber\\
&&\times\left[\frac{1}{l_+}\left\{\delta(\omega-k_++l_+)-\delta(\omega-k_+)\right\}
\right.\nonumber\\
&&\left.
+\frac{1}{q_+}\left\{\delta(\omega-k_+-q_+)-\delta(\omega-k_+)\right\}\right]\nonumber\\
(A34)&=&-ig_s^3(1-\alpha)\bar{v}\dirac{n}_\pm\Gamma T^a u\epsilon_+\nonumber\\
&&\times\int\frac{d^4l}{(2\pi)^4}\left[C_F\frac{1}{q_+}\left(\delta(\omega-k_+-q_+)+\delta(\omega-k_++l_+)\right.
\right.\nonumber\\
&&\left.
- \delta(\omega-k_+-q_++l_+)-\delta(\omega-k_+)\right)\nonumber\\
&&- \frac{C_A}{2}\left(\frac{1}{l_++q_+}\left\{\delta(\omega-k_+-q_+)-\delta(\omega-k_++l_+)\right\}
\right.\nonumber\\
&&\left.
- \left.\frac{1}{q_+}\left\{\delta(\omega-k_+-q_++l_+)-\delta(\omega-k_++l_+)\right\}\right)\right]\nonumber
\end{eqnarray}
\begin{eqnarray}
(A44)&=&ig_s^3(1-\alpha)\bar{v}\dirac{n}_\pm\Gamma T^a u\epsilon_+\nonumber\\\nonumber\\
&&\times\dint{4}{l}\frac{1}{l^4}\left[C_F\frac{1}{q_+}\Bigg\{\delta(\omega-k_++l_+)-\delta(\omega-k_+)\right.\nonumber\\
&&+ \delta(\omega-k_+-q_+)-\delta(\omega-k_+-q_++l_+)
\left. 
- \frac{l_+}{q_+}\left(\delta(\omega-k_+-q_+)-\delta(\omega-k_+)\right)\right\}\nonumber\\
&&+ \frac{C_A}{2}\frac{l_+}{q_+}\left\{\frac{1}{l_++q_+}\left(\delta(\omega-k_+-q_+)-\delta(\omega-k_++l_+)\right)
\right.\nonumber\\
&&\left.
+\left.\frac{1}{l_+-q_+}\left(\delta(\omega-k_+-q_++l_+)-\delta(\omega-k_+)\right)\right\}\right]\nonumber
\\
(B12)&=&\frac{3C_A}{4}\frac{\alpha_s}{4\pi}g_s (1-\alpha)\frac{1}{\varepsilon} \delta(\omega-k_+)\bar{v}\dirac{n}_\pm\Gamma T^a u \frac{v\cdot\epsilon}{v\cdot q}\nonumber\\
(B13)&=&0\nonumber\\
(B14)&=&0\nonumber\\
(B15)&=&-\frac{1}{2}(C_A-2C_F)\frac{\alpha_s}{4\pi} g_s (1-\alpha)\frac{1}{\varepsilon} \delta(\omega-k_+)\bar{v}\dirac{n}_\pm\Gamma T^a u \frac{v\cdot\epsilon}{v\cdot q}\nonumber\\
(B23)&=&-ig_s^3(1-\alpha)\frac{C_A}{4}\int\frac{d^4l}{(2\pi)^4}\delta(\omega-k_++l_+)\nonumber\\
&&\times\left[\frac{1}{q_+}\left(\frac{1}{(l+q)^4}-\frac{1}{l^4}\right)\bar{v}\dirac{n}_\pm\Gamma T^a u\epsilon_+\right.\nonumber\\
&& \left.+\frac{1}{k_++q_+}\left(\frac{1}{(k-l)^2(l+q)^2}-\frac{1}{(l+q)^4}\right)\right]\bar{v}\dirac{\epsilon}\dirac{n}_+\dirac{n}_\pm\Gamma T^a u\nonumber\\
(B34)&=&ig_s^3(1-\alpha)C_F\dint{4}{l}\frac{1}{l^4}\left\{\delta(\omega-k_++l_+)-\delta(\omega-k_+)\right\}
\bar{v}\dirac{n}_\pm\Gamma T^a u\frac{v\cdot \epsilon}{v\cdot q}\nonumber\\
(B35)&=&-ig_s^3(1-\alpha)C_F\dint{4}{l}\frac{1}{l^4}\delta(\omega-k_++l_+)
\bar{v}\dirac{n}_\pm\Gamma T^a u\frac{v\cdot \epsilon}{v\cdot q}\nonumber\\
(B44)&=&-ig_s^3(1-\alpha)C_F\int \frac{d^4l}{(2\pi)^4}\frac{1}{l^4}\left\{\delta(\omega-k_++l_+)-\delta(\omega-k_+)\right\}
\bar{v}\dirac{n}_\pm\Gamma T^a u \frac{v\cdot\epsilon}{v\cdot q}\nonumber\\
(B45)&=&ig_s^3(1-\alpha)C_F\dint{4}{l}\frac{1}{l^4}\left\{\delta(\omega-k_+-l_+)-\delta(\omega-k_+)\right\}
\bar{v}\dirac{n}_\pm\Gamma T^a u\frac{v\cdot \epsilon}{v\cdot q}\nonumber\\
(B55)&=&-C_F\frac{\alpha_s}{4\pi} g_s(1-\alpha)\frac{1}{\varepsilon}\delta(\omega-k_+)\bar{v}\dirac{n}_\pm\Gamma T^a u \frac{v\cdot\epsilon}{v\cdot q}\nonumber\\
(C12)&=&-\frac{3C_A}{4}\frac{\alpha_s}{4\pi}g_s (1-\alpha)\frac{1}{\varepsilon}\delta(\omega-k_+-q_+) \frac{1}{(k+q)^2}\bar{v}\dirac{\epsilon}(\dirac{k}+\dirac{q})\dirac{n}_\pm\Gamma T^a u\nonumber\\
(C13)&=&\frac{i}{2}g_s^3(1-\alpha)(C_A-2C_F)\bar{v}\dirac{\epsilon}\dirac{n}_+\dirac{n}_\pm\Gamma T^a u\nonumber\\
&&\times\int\frac{d^4l}{(2\pi)^4}\frac{1}{k_++q_+}\delta(\omega-k_+-q_++l_+)\left[\frac{1}{l^2(k+q-l)^2}-\frac{1}{l^4}\right]\nonumber\\
(C14)&=&\frac{i}{2}g_s^3(1-\alpha)(C_A-2C_F)\int \frac{d^4}{(2\pi)^4}\frac{1}{l^2(k+q-l)^2}\frac{1}{k_++q_+}\nonumber\\
&&\times\left\{\delta(\omega-k_+-q_+)-\delta(\omega-k_+-q_++l_+)\right\}\bar{v}\dirac{\epsilon}\dirac{n}_+\dirac{n}_\pm\Gamma T^a u\nonumber\\
(C15)&=&\frac{1}{2}(C_A-2C_F)\frac{\alpha_s}{4\pi}g_s (1-\alpha)\frac{1}{\varepsilon}\delta(\omega-k_+-q_+) \frac{1}{(k+q)^2}\bar{v}\dirac{\epsilon}(\dirac{k}+\dirac{q})\dirac{n}_\pm\Gamma T^a u\nonumber\\
(C34)&=&-ig_s^3(1-\alpha)C_F\dint{4}{l}\frac{1}{l^4}\frac{1}{(k+q)^2}\bar{v}\dirac{\epsilon}(\dirac{k}+\dirac{q})\dirac{n}_\pm\Gamma T^a u\nonumber\\
&&\times\left\{\delta(\omega-k_+-q_+-l_+)-\delta(\omega-k_+-q_+)\right\}\nonumber
\end{eqnarray}
\begin{eqnarray}
(C35)&=&ig_s^3(1-\alpha)C_F\dint{4}{l}\left[\frac{1}{l^4}\frac{1}{(k+q)^2}\bar{v}\dirac{\epsilon}(\dirac{k}+\dirac{q})\dirac{n}_\pm\Gamma T^a u\right.\nonumber\\
&&+\left.\frac{1}{2}\left[\frac{1}{l^2(k+q-l)^2}-\frac{1}{l^4}\right]\frac{1}{k_++q_+}\bar{v}\dirac{\epsilon}\dirac{n}_+\dirac{n}_\pm\Gamma T^a u\right]
\delta(\omega-k_+-q_++l_+)\nonumber\\
(C44)&=&ig_s^3(1-\alpha)C_F\int\frac{d^4l}{(2\pi)^4}\frac{1}{l^4}\left\{\delta(\omega-k_++l_+)-\delta(\omega-k_+)\right\}
\bar{v}\dirac{\epsilon}(\dirac{k}+\dirac{q})\dirac{n}_\pm\Gamma T^a u\frac{1}{(k+q)^2}\nonumber\\
(C45)&=&-ig_s^3(1-\alpha)C_F\dint{4}{l}\left[\frac{1}{l^4}\frac{1}{(k+q)^2}\bar{v}\dirac{\epsilon}(\dirac{k}+\dirac{q})\dirac{n}_\pm\Gamma T^a u\right.\nonumber\\
&&+\left.\frac{1}{2}\left[\frac{1}{l^2(k+q-l)^2}-\frac{1}{l^4}\right]\frac{1}{k_++q_+}\bar{v}\dirac{\epsilon}\dirac{n}_+\dirac{n}_\pm\Gamma T^a u\right]\nonumber\\
&&\times\left\{\delta(\omega-k_+-q_++l_+)-\delta(\omega-k_+-q_+)\right\}\nonumber\\
(C55)&=&C_F\frac{\alpha_s}{4\pi}g_s(1-\alpha)\frac{1}{\varepsilon}\delta(\omega-k_+-q_+) \frac{1}{(k+q)^2}\bar{v}\dirac{\epsilon}(\dirac{k}+\dirac{q})\dirac{n}_\pm\Gamma T^a u\nonumber
\label{gauge-form}
\end{eqnarray}
As mentioned at the end of Sec.~6, the terms proportional to $\dirac{\epsilon}\dirac{n}_+\dirac{n}_\pm$ vanish for $\phi_+$. In addition, all the integrals must be understood with an infrared regulator, since we are only interested in their ultraviolet behaviour.
\end{appendix}

\end{document}